\documentclass[a4paper,11pt]{article}
\pdfoutput=1 

\usepackage{jcappub} 

\usepackage{dcolumn}
\usepackage{bm}
\usepackage{physics}
\usepackage{xcolor}
\usepackage{comment}
\usepackage{amsmath}
\usepackage{mathrsfs}

\newcommand{\mpl}{M_{\mathrm{pl}}}

\newcommand{\bz}{\bm{\zeta}}
\newcommand{\bc}{\bm{\chi}}
\newcommand{\md}{\mathrm{d}}

\newcommand{\bx}{\bm{x}}

\newcommand{\bp}{\bm{p}}
\newcommand{\bq}{\bm{q}}
\newcommand{\bk}{\bm{k}}

\title{\boldmath Generating large primordial fluctuations in single-field inflation for PBH formation}

\author[1,2]{Jason Kristiano}
\author[3,1,2,4]{and Jun'ichi Yokoyama}

\affiliation[1]{Research Center for the Early Universe (RESCEU), Graduate School of Science, The University of Tokyo, Tokyo 113-0033, Japan}
\affiliation[2]{Department of Physics, Graduate School of Science, The University of Tokyo, Tokyo 113-0033, Japan}
\affiliation[3]{Kavli Institute for the Physics and Mathematics of the Universe (Kavli IPMU), WPI, UTIAS, The University of Tokyo, Kashiwa, Chiba 277-8568, Japan}
\affiliation[4]{Trans-Scale Quantum Science Institute, The University of Tokyo, Tokyo 113-0033, Japan}

\emailAdd{jkristiano@resceu.s.u-tokyo.ac.jp}
\emailAdd{yokoyama@resceu.s.u-tokyo.ac.jp}

\abstract{
In order to produce appreciable amount of primordial black holes (PBHs), the square amplitude of curvature perturbation must take a large value of $\mathcal{O}(0.01)$, namely, seven digits larger than the value observed by cosmic microwave background radiation (CMB) on large scales. Such a large fluctuation can be achieved by violating the slow-roll (SR) condition within a short duration.  The best known of such possibilities is the ultraslow-roll (USR) inflation. We calculate the power spectrum of curvature perturbation in a simple single-field inflation model which evolves through the SR-USR-SR regimes so that both large-amplitude small-scale fluctuation for PBH formation and small-amplitude large-scale fluctuation as observed by CMB are realized. We further calculate the bispectrum and one-loop correction to the power spectrum induced by the third-order action of curvature perturbation as the beginning of precision cosmology on small scales. As a result, we show that single-field inflation model realizing PBH formation can be constrained by the quantum correction.
}

\begin{document}
\maketitle
\flushbottom

\section{Introduction}
PBHs are formed when an overdense region with a magnitude of density fluctuation close to unity enters the Hubble horizon during radiation domination \cite{Zel:1967, Hawking:1971ei, Carr:1974nx, Hawking:1974rv}.
The most popular formation mechanism of such a large perturbation on small scales relevant to PBHs is quantum effects during cosmic inflation in the very early universe \cite{Starobinsky:1980te,Sato:1980yn,Guth:1980zm} (see \cite{Sato:2015dga} for a review). In this chapter, we focus on single-field inflation models that can produce large fluctuations on small scales relevant to PBH formation with particular attention to our recent finding \cite{Kristiano:2022maq, Kristiano:2023scm} that higher-order quantum effects can restrict the feasibility of such models, even though they are observationally possible because there are no stringent observational constraints on primordial fluctuations on small scales \cite{Chluba:2012we, Chluba:2019nxa, Nakama:2014vla, Jeong:2014gna, Inomata:2016uip, Nakama:2017qac, Kawasaki:2021yek, Kimura:2021sqz, Wang:2022nml} unlike on large scales where observations of CMB tightly constrain the fluctuation spectrum \cite{Planck:2018nkj, Planck:2018jri, Planck:2019kim}.

\begin{figure}[tbp]
\centering 
\includegraphics[width=0.8\textwidth]{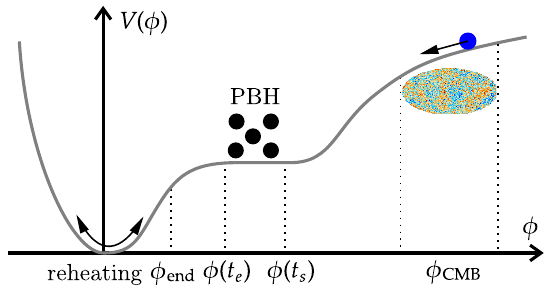}
\caption{\label{fig1} Schematic picture of the inflaton potential realizing PBH formation from USR condition. When the inflaton is around $\phi_\mathrm{CMB}$, scales probed by CMB observations leave the horizon and it is in the SR regime. It enters an extremely flat region at $t=t_s$ undergoing an USR period. It enters the SR period again at $t=t_e$ until $\phi_\mathrm{end}$, the end of inflation.}
\end{figure}

The simplest inflation model consistent with CMB observation is canonical SR inflation, where the scalar field which drives the inflation, called inflaton, is described by a canonical kinetic term and a potential. At a certain range, the potential has to be slightly tilted to realize the SR condition, so it can explain CMB observations. Beyond such a range, one can construct a theory that violates the SR condition, which usually results in the amplification of fluctuation on small scales \cite{Yokoyama:1998rw, Saito:2008em} due to rapid slowdown of the inflaton's velocity. The simplest example of a violation of the SR condition is the USR condition \cite{Kinney:1997ne, Inoue:2001zt, Kinney:2005vj, Martin:2012pe, Motohashi:2017kbs}, which can be generalized to the constant-roll condition \cite{Motohashi:2014ppa, Motohashi:2017aob, Motohashi:2019rhu}. It occurs when the inflaton enters an extremely flat region in the potential \cite{Ivanov:1994pa}, where the first derivative of the potential vanishes. Peaks of the enhanced fluctuation can collapse into PBH after such a fluctuation reenters the horizon. An example of potential that realizes the formation of PBHs while satisfying the observational constraint on large scales is shown in Fig.~\ref{fig1}.

In this chapter, we explain how violation of the SR approximation can lead to amplification of primordial fluctuation on certain scales. Although we choose to focus on USR inflation, the method presented in this chapter can be easily generalized to other mechanism that yields rapid slowdown of the inflaton's velocity. In Sec.~\ref{sec2}, we introduce two-point functions or the power spectrum of the fluctuations in SR inflation, USR inflation, and USR to SR transition. In Sec.~\ref{sec3}, we introduce three-point functions of fluctuations in the same setup as in Sec.~\ref{sec2}. In Sec.~\ref{sec4}, we work out an inflation model with SR-USR-SR transitions. In this model, fluctuations are amplified on scales corresponding to the time of the USR phase. In Sec.~\ref{sec5}, we discuss other PBH formation mechanisms within a single-field inflation framework and summarize our review.

\section{Two-point functions \label{sec2}}
The action of canonical inflation is given by
\begin{equation}
S = \frac{1}{2} \int \md^4x \sqrt{-g} \left[ \mpl^2 R - (\partial_\mu \phi)^2 - 2 V(\phi) \right], \label{action}
\end{equation}
where $\mpl$ is reduced Planck scale, $g = \mathrm{det}~g_{\mu\nu}$, $g_{\mu\nu}$ and $R$ are metric tensor and its Ricci scalar. Consider a spatially flat, homogeneous, and isotropic background,
\begin{equation}
\md s^2 = -\md t^2 + a^2(t) \md \mathbf{x}^2 = a^2(\tau) (-\md \tau^2 + \md \mathbf{x}^2),
\end{equation}
where $\tau$ is conformal time. Equations of motion for the scale factor $a(t)$ and the homogeneous part of the inflaton $\phi(t)$ are the Friedmann equations
\begin{equation}
H^2 = \frac{1}{3\mpl^2}\left(\frac{1}{2} \dot{\phi}^2 + V(\phi)\right), ~\dot{H} = - \frac{\dot{\phi}^2}{2 \mpl^2},
\end{equation}
with $H=\dot{a}/{a}$ being the Hubble parameter, and the Klein-Gordon equation
\begin{equation}
\ddot{\phi} + 3 H \dot{\phi} + \frac{\md V}{\md\phi} = 0. \label{klein}
\end{equation}
Here, a dot denotes time derivative.

During inflation, the evolution of the Hubble parameter is parameterized by the slow-roll (SR) parameters. The SR parameters are defined as
\begin{equation}
\epsilon_{n+1} = \frac{\dot{\epsilon}_n}{\epsilon_n H},  ~ \epsilon_1 \equiv - \frac{\dot{H}}{H^2} , 
\end{equation}
where $n$ is a positive integer. In case the Hubble parameter $H$ is almost a constant, we have $\epsilon_1 \ll 1$. This condition is called the quasi-de Sitter (quasi-dS) approximation. Thus, the scale factor can be approximated as
\begin{equation}
a \approx -\frac{1}{H \tau} \propto e^{H t}.
\end{equation}
Another stronger condition is called the SR approximation, which is defined as $| \epsilon_n | \ll 1$ for every $n$. Therefore, the SR approximation implies a quasi-dS approximation, while the converse statement is not true. Violation of the SR approximation does not imply a breakdown of the quasi-dS approximation.

Small perturbation from the homogeneous part, $\phi(t)$, of the inflaton $\phi(\bx, t)$ and metric can be expressed as 
\begin{gather}
\phi(\bx,t) = \phi(t) + \delta \phi(\bx,t), \nonumber\\
\md s^2 = -N^2 \mathrm{d}t^2 + \gamma_{ij} (\mathrm{d}x^i + N^i \mathrm{d}t)(\mathrm{d}x^j + N^j \mathrm{d}t),
\end{gather}
where $\gamma_{ij}$ is the three-dimensional metric on slices of constant $t$, $N$ is the lapse function, and $N^i$ is the shift vector. We choose comoving gauge condition
\begin{equation}
\delta \phi(\bx,t) = 0, ~\gamma_{ij}(\bx,t) = a^2(t) e^{2\zeta(\bx,t)} \delta_{ij},
\end{equation}
where $\zeta(\bx,t)$ is comoving curvature perturbation. $N$ and $N^i$ are obtained by solving the constraint equations. In this review, we do not consider the tensor perturbation.

Expanding the action \eqref{action} up to the second order of the curvature perturbation yields
\begin{equation}
S^{(2)}[\zeta] = M_{\mathrm{pl}}^2 \int \mathrm{d}t ~\mathrm{d}^3x ~a^3 \epsilon_1  \left[ \dot{\zeta}^2 - \frac{1}{a^2} (\partial_i \zeta)^2  \right].
\label{S2}
\end{equation}
Introducing the Mukhanov-Sasaki variable by $v = z \mpl \zeta$, where $z = a \sqrt{2\epsilon_1}$, the action becomes canonically normalized
\begin{equation}
S^{(2)}_{\mathrm{can}} [v] = \frac{1}{2} \int \mathrm{d}\tau ~\mathrm{d}^3x \left[ (v')^2 - (\partial_i v)^2 + \frac{z''}{z} v^2 \right], \label{action2}
\end{equation}
where a prime denotes derivative with respect to $\tau$. We can see that the quantity $z''/z$ behaves as an effective time-dependent mass squared. In momentum space, quantization is performed by promoting the Mukhanov-Sasaki variable to an operator
\begin{equation}
\hat{v}_{\bk} (\tau) = \mpl z(\tau) \hat{\zeta}_{\bk} (\tau) =  v_k(\tau) \hat{a}_{\bk} + v^*_k (\tau) \hat{a}_{-\bk}^\dagger, \label{voperator}
\end{equation}
where mode function $v_k(\tau)$ satisfies the Mukhanov-Sasaki equation
\begin{equation}
v_k'' + \left( k^2 - \frac{z''}{z} \right) v_k = 0.
\end{equation}
In terms of the SR parameters, the effective mass term can be written as
\begin{equation}
\frac{z''}{z} = (a H)^2 \left( 2 - \epsilon_1 +\frac{3}{2} \epsilon_2 -\frac{1}{2} \epsilon_1 \epsilon_2 + \frac{1}{4} \epsilon_2^2 + \frac{1}{2} \epsilon_2 \epsilon_3 \right).
\end{equation}

From (\ref{action2}) we find that $\pi^v = v'$ is momentum conjugate to $v$, so canonical quantization imposes a commutator
\begin{equation}
[\hat{v}_{\bk} (\tau),\hat{\pi}_{\bk'}^v (\tau)] = i \hbar \delta(\bk-\bk'), \label{quantization}
\end{equation}
and the operators satisfy  $[ \hat{a}_{\bk}, \hat{a}_{\bk'}^\dagger ] = (2 \pi)^3 \delta(\bk - \bk')$ under the normalization condition
\begin{equation}
v_k'^* v_k - v_k' v_k^* = i \hbar. \label{normalization}
\end{equation}
The two-point functions of curvature perturbation and power spectrum can be written as
\begin{gather}
\left\langle \zeta_{\bk} (\tau) \zeta_{\bk'} (\tau) \right\rangle \equiv (2 \pi)^3 \delta^3(\bk + \bk') \left\langle \! \left\langle \zeta_{\bk} (\tau) \zeta_{-\bk} (\tau) \right\rangle \! \right\rangle, \\
\Delta^2_s(k, \tau) \equiv \frac{k^3}{2 \pi^2} \langle \! \langle \zeta_{\bk} (\tau) \zeta_{-\bk} (\tau) \rangle \! \rangle, 
\end{gather}
the bracket $\langle \cdots \rangle = \bra{0} \cdots \ket{0}$ denotes the vacuum expectation value (VEV), and $\Delta^2_s(k, \tau)$ is the power spectrum multiplied by the phase space density.

\subsection{Slow-roll approximation}
With SR approximation, the effective mass becomes
\begin{equation}
\frac{z''}{z} = (a H)^2 \left[ 2 - \epsilon_1 +\frac{3}{2} \epsilon_2 + \mathcal{O}(\epsilon^2) \right].
\end{equation}
Up to first-order correction in SR approximation, $aH$ as function of $\tau$ reads
\begin{equation}
aH = - \frac{1}{\tau ( 1-\epsilon_1)}. \label{aht}
\end{equation}
Then, the effective mass can be expressed in terms of the conformal time as
\begin{equation}
\frac{z''}{z} = \frac{1}{\tau^2} \left( 2 + 3\epsilon_1 + \frac{3}{2} \epsilon_2 \right) = \frac{1}{\tau^2} \left( \nu^2 - \frac{1}{4} \right),
\end{equation}
where $\nu$ is defined as
\begin{equation}
\nu = \frac{3}{2} + \delta = \frac{3}{2} + \epsilon_1 + \frac{1}{2} \epsilon_2 .
\end{equation}

At early time, $\tau \rightarrow -\infty$, the Mukhanov-Sasaki equation approaches
\begin{equation}
v_k'' + k^2 v_k = 0,
\end{equation}
which is the equation of a simple harmonic oscillator. Requiring minimum energy state and normalization \eqref{normalization}, solution of the Mukhanov-Sasaki equation at early time is given by
\begin{equation}
v_k (\tau \rightarrow -\infty) = \frac{1}{\sqrt{2k}} e^{-i k\tau}. \label{vearly}
\end{equation}
This minimum energy state corresponds to the initial vacuum state called Bunch-Davies vacuum, defined as $\hat{a}_{\bk} \ket{0} = 0$.

At general time, solution of the Mukhanov-Sasaki equation is given by
\begin{equation}
v_k(\tau) = \sqrt{- k \tau} \left[ c_1 H_\nu^{(1)}(- k\tau) + c_2 H_\nu^{(2)}(- k\tau) \right], \label{mshankel}
\end{equation}
where $c_1$ and $c_2$ are coefficients to be fixed to a boundary condition. Given the asymptotic limit of Hankel functions
\begin{equation}
\lim_{x \rightarrow \infty} H_\nu^{(1,2)}(x) = \sqrt{ \frac{2}{\pi x} } \exp\left[ \pm i \left( x - \frac{\nu \pi}{2} - \frac{\pi}{4} \right) \right],
\end{equation}
we can match the general solution to the early time limit \eqref{vearly} that leads to
\begin{equation}
c_1 = \sqrt{ \frac{\pi}{4 k} } e^{i \frac{\pi}{4} (2\nu +1)} ~~\mathrm{and}~~ c_2 = 0.
\end{equation}
Then, the mode function of Mukhanov-Sasaki variable becomes
\begin{equation}
\label{vkhankel}
v_k(\tau) = \sqrt{ \frac{\pi}{4}} e^{i \frac{\pi}{4} (2\nu +1)} \sqrt{- \tau} H_\nu^{(1)}(- k\tau).
\end{equation}

To obtain the mode function of the curvature perturbation, we need to know the explicit function of $z(\tau)$. Starting from
\begin{equation}
\frac{z'}{z} = aH \left( 1 + \frac{1}{2} \epsilon_2 \right),
\end{equation}
we can substitute \eqref{aht} to obtain
\begin{align}
\int \frac{\md z}{z} &= - \int \frac{\md\tau}{\tau} \left( 1 + \epsilon_1 + \frac{1}{2} \epsilon_2 \right) \nonumber\\
\frac{z(\tau)}{z_\star} &= \left( \frac{\tau}{\tau_\star}  \right)^{- \left( 1 + \epsilon_1 + \frac{1}{2} \epsilon_2 \right) }, 
\end{align}
where $\tau_\star$ is an arbitrary time reference and $z_\star = z(\tau_\star)$. Choosing $\tau_\star$ as horizon crossing time $k = a(\tau_\star) H(\tau_\star)$ so
\begin{align}
\tau_\star &= - \left[ \frac{1}{aH (1-\epsilon_1)} \right]_\star = - \frac{1}{ k (1-\epsilon_1)_\star}, \label{tauh} \\
z_\star &= \left( a \sqrt{2\epsilon_1} \right)_\star = \left( \frac{k}{ H} \sqrt{2\epsilon_1} \right)_\star,
\end{align}
then the explicit function of $z(\tau)$ is
\begin{equation}
\label{ztau}
z(\tau) = \left( \frac{k}{ H} \sqrt{2\epsilon_1} \right)_\star \left[ - k\tau (1-\epsilon_1)_\star \right]^{-\left( 1 + \epsilon_1 + \frac{1}{2} \epsilon_2  \right)}.
\end{equation}

Then, the mode function of curvature perturbation reads
\begin{equation}
\zeta_k(\tau) = \frac{v_k(\tau)}{\mpl z(\tau)} = \left( \frac{H}{\mpl k \sqrt{2 \epsilon_1}} \right)_\star \sqrt{ \frac{\pi }{4}  } (- k \tau)^{\left( 1 + \epsilon_1 + \frac{1}{2} \epsilon_2  \right)}   e^{i \frac{\pi}{4} (2\nu +1)}  \sqrt{-\tau } H_\nu^{(1)}(-k\tau). \label{zeta}
\end{equation}
With another asymptotic limit of Hankel function
\begin{equation}
\label{limhan}
\lim_{x \rightarrow 0} H_\nu^{(1)}(x) = - \frac{i}{\pi} \Gamma(\nu) \left( \frac{x}{2} \right)^{-\nu},
\end{equation}
the curvature perturbation at the end of inflation $\tau = \tau_0 (\rightarrow 0)$, approaches
\begin{equation}
\zeta_k(\tau_0) = \left(  \frac{i H}{\mpl  \sqrt{2 k^3 \epsilon_1}} \right)_\star \frac{ 1 }{ \sqrt{4 \pi} } 2^{3/2} \Gamma(3/2)  . 
\end{equation}
Also, the power spectrum reads
\begin{equation}
 \Delta^2_{s}(k, \tau_0) = \left( \frac{H^2}{8 \pi^2 M_{\mathrm{pl}}^2 \epsilon_1} \right)_\star, \label{pssr}
\end{equation}
with a weak $k$-dependence due to the horizon crossing condition manifested in the spectral tilt
\begin{equation}
n_s (k, \tau_0) - 1 = \frac{\md \log \Delta_{s}^{2}(k, \tau_0)}{\md \log k} = (- 2 \epsilon_1 - \epsilon_2)_\star ~. \label{tiltsr}
\end{equation}

Another way to derive power spectrum in SR approximation is by looking at exact de Sitter case. In exact de Sitter, the effective mass is
\begin{equation}
\frac{z''}{z} = \frac{2}{\tau^2} .
\end{equation}
Solution of the Mukhanov-Sasaki equation with Bunch-Davies initial vacuum is
\begin{equation}
v_k(\tau) = \frac{1}{\sqrt{2k}} \left( 1-\frac{i}{k\tau} \right) e^{-ik\tau} . \label{msds}
\end{equation}
When we proceed to derive the curvature perturbation, we run into a trouble because in exact de Sitter we find $\epsilon_1 = 0$. However, as a first-order approximation of the SR parameters, we can simply consider the finite value of $\epsilon_1$. Also, the Hubble and SR parameters have a weak time dependence. 

Numerical calculation of $\zeta_k(\tau)$ shows that evaluating the proportionality coefficient $1/z$ to $v_k(\tau)$ at the horizon crossing time
$\tau = \tau_\star$ yields an excellent fit.
So the curvature perturbation \eqref{zeta} can be approximated as
\begin{equation}
\zeta_k(\tau) = \left( \frac{i H}{2 \mpl \sqrt{k^3 \epsilon_1}} \right)_\star (1+ik\tau) e^{-ik\tau}.
\end{equation}
Substituting $\tau \rightarrow 0$ leads to the power spectrum \eqref{pssr}.

The power spectrum \eqref{pssr} can be expressed as a function of potential. We can rewrite the Friedmann equation as
\begin{equation}
V = H^2 \mpl^2 (3 - \epsilon_1) .
\end{equation}
Taking derivative with respect to $\phi$ yields
\begin{equation}
V_{,\phi} \equiv \frac{\md V }{\md \phi} = \frac{1}{\sqrt{2}} H^2 \mpl (6 - 2 \epsilon_1 + \epsilon_2).
\end{equation}
At leading order in  the SR approximation, the first SR parameter can be written as
\begin{equation}
\epsilon_1 \simeq \frac{\mpl^2}{2} \left( \frac{V_{,\phi}}{V} \right)^2.
\end{equation}
Therefore, $\epsilon_1$ has an implicit time dependence because $\phi$ depends on time.

From \eqref{pssr}, we can see that the power spectrum at the wavenumber $p$ corresponds to $\epsilon_1$ evaluated at $\tau = -1/p$. This implies that observation of the power spectrum at the wavenumber $p$ constrains $V_{,\phi}(\phi(\tau))$ at $\tau = -1/p$. To solve the horizon problem, the fluctuations in the CMB must be around $60$ e-folds before the end of inflation. Moreover, fluctuations are observed within the finite range of wavenumber $0.005 \mathrm{Mpc}^{-1} < p < 0.2 \mathrm{Mpc}^{-1}$ by CMB. This means that CMB observations constrain only a finite region of the potential around $60$ e-folds before the end of inflation. Beyond such a region, there are much looser observational constraints on the potential. Therefore, it is possible to have an inflation model in which the SR approximation is violated after the horizon-crossing time of the CMB scales and before the end of inflation. 

Comparison with observations can only be made if and only if we are convinced that quantum fluctuations generated during inflation behave as classical statistical fluctuations to leave observable traces in our universe. The following intuitive argument can be developed for the classicalization of curvature fluctuations.
In superhorizon regime, $k\ll a H$, the mode function behaves as
\begin{equation}
\zeta_k(\tau) = \left( \frac{iH}{2\mpl\sqrt{ k^3 \epsilon_1}} \right)_\star \left[1+ \mathcal{O} \left(\frac{k^2}{ a^2 H ^2}\right)\right],
\end{equation}
which means $\zeta_k^{\ast}(\tau)=-\zeta_k (\tau)$ to the leading order.  So we find $\hat{\zeta}_{\bk} (\tau) = \zeta_k (\tau)(\hat{a}_{\bk}-\hat{a}_{-\bk}^{\dagger})$ and its conjugate
momentum $\hat{\pi}_{\bk}^\zeta (\tau)= \mpl^2 z^2(\tau) {\zeta}_{\bk}'(\tau)(\hat{a}_{\bk}-\hat{a}_{-\bk}^\dagger)$ have the same operator dependence, and apparently commute with each other, behaving classically \cite{Polarski:1995jg}. This is why long-wave quantum fluctuations behave as if they are classical statistical fluctuations to provide the origin of large-scale structures and CMB anisotropy.  The more precise statement, however, is that the quantum commutator
\begin{equation}
\left[ \hat{\zeta}_{\bk} (\tau),\hat{\pi}_{\bk'}^\zeta (\tau) \right]=i\hbar \delta(\bk-\bk') \label{qcom}
\end{equation}
always holds, and it is the commutator
\begin{equation}
\left[ \hat{\zeta}_{\bp} (\tau),\hat{\zeta}_{\bq}' (\tau) \right]=\frac{i\hbar}{2 \mpl^2 \epsilon_1(\tau) a^2(\tau) } \delta(\bp - \bq) \label{zetacom}
\end{equation}
that decreases exponentially during the standard slow-roll inflation.

Observation of the spectral tilt $n_s=0.9649 \pm 0.0042$ on the CMB scale implies that the SR parameters $\epsilon_1, \abs{\epsilon_2}$ at $\tau = - 1/p$ have a value $\mathcal{O}(0.01)$, based on \eqref{tiltsr}. If the first SR parameter decreases rapidly at a later time $\tau \gg -1/p$, the power spectrum is amplified at wavenumber $k \gg p$. This can be achieved by violating the SR approximation at a later time. In the next subsections, we will show some inflation models that violate the SR approximation.

\subsection{Ultraslow-roll inflation \label{usr}}
In this subsection, we consider a regime where the potential is constant, called ultra-low-roll inflation (USR) \cite{Kinney:1997ne, Inoue:2001zt, Kinney:2005vj, Martin:2012pe, Motohashi:2017kbs}. Because $\md V / \md \phi = 0$, the Klein-Gordon equation \eqref{klein} becomes $\ddot{\phi} = -3H \dot{\phi}$, so $\dot{\phi} \propto a^{-3}$. This makes $\epsilon_1$ strongly time-dependent and extremely small as
\begin{equation}
\epsilon_1 = \frac{\dot{\phi}^2}{2 \mpl^2 H^2}  \propto a^{-6}, \label{epsusr}
\end{equation}
and the second SR parameter becomes
\begin{equation}
\epsilon_2 \equiv \frac{\dot{\epsilon}_1}{\epsilon_1 H} = 2 \epsilon_1 +  2 \frac{\ddot{\phi}}{\dot{\phi} H} \simeq -6,
\end{equation}
which breaks SR approximation. The effective mass can be written as
\begin{equation}
\frac{z''}{z} = (a H)^2 \left[ 2 +\frac{3}{2} \epsilon_2 + \frac{1}{4} \epsilon_2^2 + \mathcal{O}(\epsilon_1) \right] = \frac{2}{\tau^2} ,
\end{equation}
which is the same as in the exact de Sitter case. Thus, solution of the Mukhanov-Sasaki equation with Bunch-Davies initial vacuum is the same as \eqref{msds}. Then, the curvature perturbation is given by
\begin{equation}
\zeta_k(\tau) =  \frac{i H}{2 \mpl \sqrt{k^3 \epsilon_1(\tau)}}  (1+ik\tau) e^{-ik\tau}.
\end{equation}
Here, $\epsilon_1$ has a strong dependence on time, which can be expressed as
\begin{equation}
\epsilon_1(\tau) = \epsilon_1(\tau_\star) \left( \frac{a(\tau)}{a(\tau_\star)} \right)^{-6},
\end{equation}
where $\tau_\star$ is the horizon-crossing time of a characteristic wavenumber $k_\star$. Therefore, the curvature perturbation becomes
\begin{equation}
\zeta_k(\tau) =  \frac{i H}{2 \mpl \sqrt{k^3 \epsilon_1(\tau_\star)}} \left( \frac{a(\tau)}{a(\tau_\star)} \right)^3 (1+ik\tau) e^{-ik\tau}. \label{modeusr}
\end{equation}
At the end of inflation, the power spectrum is
\begin{equation}
\Delta^2_{s}(k, \tau_0) = \frac{H^2}{8 \pi^2 M_{\mathrm{pl}}^2 \epsilon_1(\tau_\star)} \left( \frac{a(\tau_0)}{a(\tau_\star)} \right)^6.
\end{equation}

To solve the horizon problem, CMB-scale fluctuations must cross the horizon at $N_\mathrm{CMB} \approx 60$ e-folds before the end of inflation. It means that power spectrum of CMB fluctuations can be written as
\begin{equation}
\Delta^2_{s}(p, \tau_0) = \frac{H^2}{8 \pi^2 M_{\mathrm{pl}}^2 \epsilon_1(\tau_\star)} e^{6 N_\mathrm{CMB}}.
\end{equation}
To explain the observed $\Delta^2_{s}(p, \tau_0) = 2.1 \times 10^{-9}$, for a reasonable value of $\epsilon_1(\tau_\star)$, an extremely small value of $H/\mpl$ is required. Because $H/\mpl$ characterize the energy scale of inflation, an extremely small value of $H/\mpl$ can reach the energy scale of the Big Bang nucleosynthesis. Hence, inflation with only USR period cannot have the required e-folds number to solve the horizon problem.

\subsection{Constant-roll inflation}
We can generalize the USR condition to $\epsilon_2 = \mathrm{constant}$ \cite{Motohashi:2014ppa, Motohashi:2017aob, Motohashi:2019rhu}. By taking reference at time $\tau_\star$, the first SR parameter can be written as
\begin{equation}
\epsilon_1(\tau) = \epsilon_1(\tau_\star) \left( \frac{a(\tau)}{a(\tau_\star)} \right)^{\epsilon_2}, \label{eps1cr}
\end{equation}
Rewriting the second SR parameter as $\epsilon_2 = -2 (3+ \alpha)$, the effective mass can be expressed as
\begin{equation}
\frac{z''}{z} = \frac{1}{\tau^2} (2 + \alpha) (1 + \alpha) = \frac{1}{\tau^2} \left( \nu^2 - \frac{1}{4} \right),
\end{equation}
where $\nu$ is defined as
\begin{equation}
\nu = \abs{ \alpha + \frac{3}{2} } . \label{crnu}
\end{equation}
Solution of the Mukhanov-Sasaki equation with Bunch-Davies initial condition is the same as \eqref{vkhankel}. The power spectrum is given by
\begin{equation}
 \Delta^2_{s}(k, \tau_0) = \frac{H^2}{8 \pi^2 M_{\mathrm{pl}}^2 \epsilon_1(\tau_0) } \frac{1}{2 \pi} 2^{2 \nu} [\Gamma(\nu)]^2 \left( \frac{k}{aH} \right)^{3 - 2\nu}.
\end{equation}
We can read that the spectral tilt is $n_s - 1 = 3 - 2\nu$. Therefore, there are two possible solutions for $\alpha$, given by
\begin{equation}
\alpha_\pm = - \frac{3}{2} \pm \frac{1}{2} (4 - n_s).
\end{equation}

However, we have to choose a solution that does not grow outside the horizon. Recall the second-order action of $\zeta$ that reads
\begin{equation}
S^{(2)}[\zeta] = M_{\mathrm{pl}}^2 \int \mathrm{d}\tau ~\mathrm{d}^3x ~a^2 \epsilon_1  \left[ (\zeta')^2 - (\partial_i \zeta)^2 \right].
\end{equation}
The equation of motion is given by
\begin{equation}
\zeta_{\bk}'' + \frac{(a^2 \epsilon_1)'}{a^2 \epsilon_1} \zeta_{\bk}' + k^2 \zeta_{\bk}  = 0 
\end{equation}
At superhorizon scale, we can neglect the last term in the equation of motion. Hence, evolution of the curvature perturbation outside the horizon is
\begin{equation}
\zeta_k (\tau_0) \rightarrow A_k + B_k \int^{\tau_0}_{\tau_\star} \frac{\md \tau}{a^2(\tau) \epsilon_1(\tau)},
\end{equation}
where $A_k$ and $B_k$ are time-independent coefficients. The second term determines whether $\zeta_k (\tau_0)$ grows or decays. Substituting \eqref{eps1cr}, the integral becomes
\begin{equation}
\int^{\tau_0}_{\tau_\star}  \frac{\md \tau}{a^2(\tau) \epsilon_1(\tau)} \sim \tau_0^{\epsilon_2 + 3} . \label{crgrow}
\end{equation}
It shows that curvature perturbation grows outside the horizon if $\epsilon_2 < -3$ or equivalently $\alpha > -3/2$. It was pointed out in \cite{Saito:2008em} that the curvature perturbation grows outside the horizon if $3 - \epsilon_1 + \epsilon_2 < 0$ is satisfied. Clearly, USR condition satisfies this condition since $\epsilon_2 = -6$. 

From observation of spectral tilt $n_s=0.9649 \pm 0.0042$ on CMB scale, we obtain $\alpha_+ \approx 0$ and $\alpha_- \approx -3$. Solution $\alpha_+ \approx 0$ corresponds to $\epsilon_2 \approx -6$, which approximately equal to USR inflation. As explained in the previous subsection, although it yields an almost scale-invariant power spectrum, this solution cannot explain the required number of e-folds of inflation because the curvature perturbation grows outside the horizon. Therefore, we have to choose $\alpha_-$ as the solution because it corresponds to decaying curvature perturbation. With this solution, constant-roll inflation can explain the observed CMB fluctuations.

\subsection{Ultraslow-roll to slow-roll transition \label{usrsr}}
\begin{figure}[tbp]
\centering 
\includegraphics[width=0.5\textwidth]{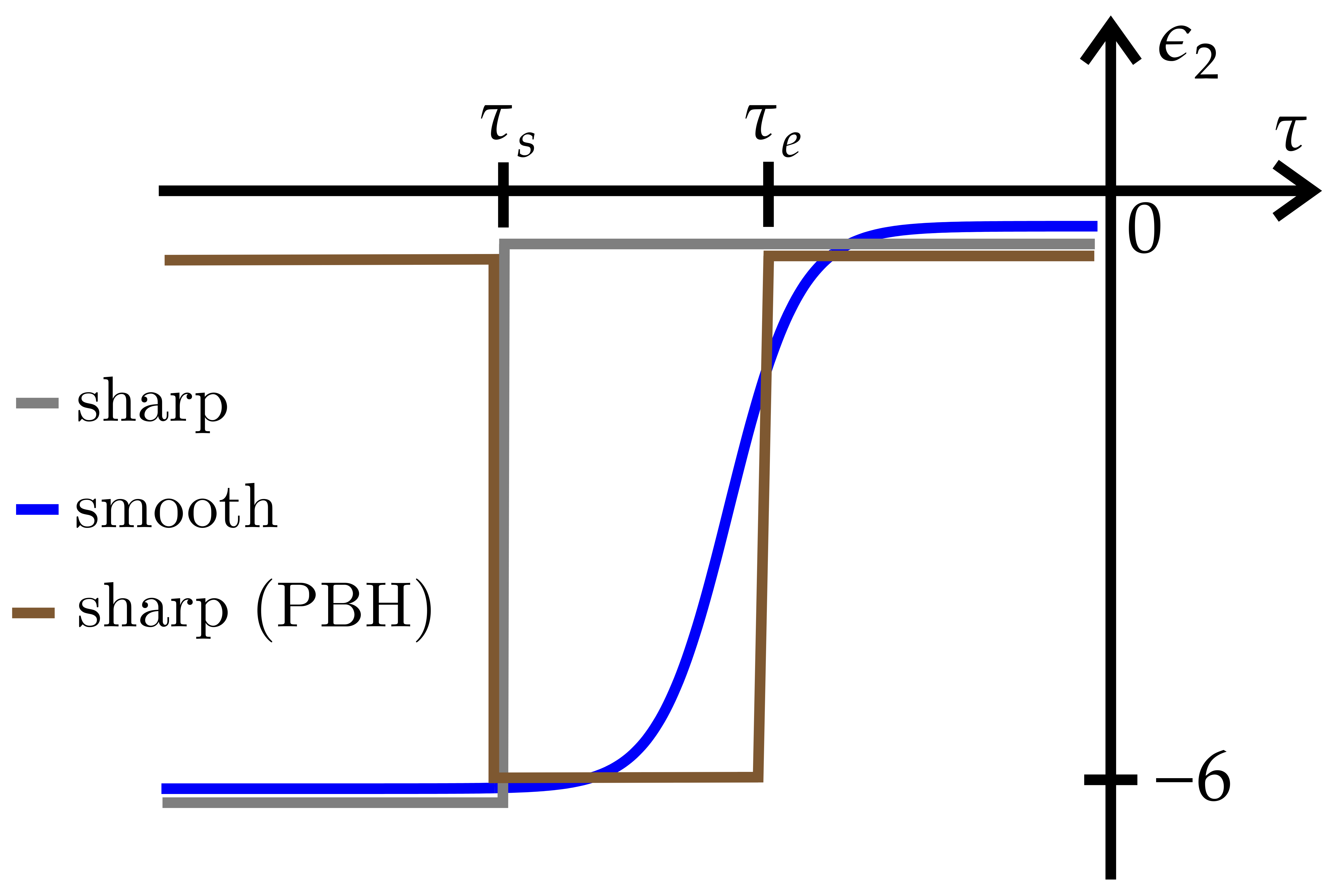}
\caption{\label{fig2} Sketch of $\epsilon_2(\tau)$ for sharp and smooth transition from USR to SR period are shown by the gray and blue curve, respectively. Brown curve shows sharp transition from SR-USR-SR period, which will be discussed in Sec.~\ref{sec4}.}
\end{figure}

In this subsection, we consider an inflation model in which the second parameter SR evolves from the USR to the SR period at conformal time $\tau_s$. For $\tau < \tau_s$, the inflaton is in the USR period. At time $\tau \geq \tau_s$, the inflaton undergoes a transition from the USR period to the SR period. In the next part, we consider two different evolutions of the second SR parameter: smooth and sharp transitions. The former means that the second SR parameter evolves as a continuous function of time, while in the latter it evolves as a step function of time. We follow the parameterization of $\epsilon_2(\tau)$ in \cite{Cai:2018dkf}, which is also used in many literatures. The sketches of $\epsilon_2(\tau)$ for both cases are shown by the gray and blue curves in Fig. \ref{fig2}.

\subsubsection{Smooth transition}
At time $\tau > \tau_s$, consider evolution of the first and second SR parameter as
\begin{gather}
\epsilon_1(\tau) = \epsilon_1(\tau_s) \frac{3 \left[ \left( \dfrac{\tau_s}{\tau} \right)^s (s-3) + s + 3 \right]^2}{2 s^2(s+3)} \left( \frac{\tau}{\tau_s} \right)^{s+3}, \label{epssmooth} \\
\epsilon_2(\tau) = s-3 + \frac{2s (s+3)}{ \left( \dfrac{\tau_s}{\tau} \right)^s (s-3) + s + 3 }, \label{etasmooth}
\end{gather}
where $s = \sqrt{9 - 12 \eta_V} \simeq 3- 2 \eta_V$ and $\eta_V \rightarrow 0$. The parameter $\eta_V$ corresponds to the second derivative of the potential at the transition point. In this model, the second SR parameter evolves from USR period $\epsilon_2(\tau_s) = - 6$ to SR period $\epsilon_2(\tau_0) = - 2 \eta_V$. With this parametrization, the effective mass becomes
\begin{equation}
\frac{z''}{z} = \frac{1}{\tau^2} \left( 2 - 3 \eta_V \right) = \frac{1}{\tau^2} \left( \nu^2 - \frac{1}{4} \right),
\end{equation}
where
\begin{equation}
\nu^2 = \frac{9}{4} - 3 \eta_V.
\end{equation}
Because $\eta_V \rightarrow 0$, solution of the Mukhanov-Sasaki equation with Bunch-Davies initial vacuum is the same as \eqref{msds}. Then, the curvature perturbation is given by
\begin{equation}
\zeta_k(\tau) =  \frac{i H}{2 \mpl \sqrt{k^3 \epsilon_1(\tau)}}  (1+ik\tau) e^{-ik\tau}, \label{zetasmooth}
\end{equation}
and power spectrum at the end of inflation becomes
\begin{equation}
\Delta^2_{s}(k, \tau_0) = \frac{H^2}{8 \pi^2 M_{\mathrm{pl}}^2 \epsilon_1(\tau_0)}.
\end{equation}

\subsubsection{Sharp transition \label{s2sharp}} 
Consider evolution of the second SR parameter as a step function of time
\begin{equation}
\epsilon_2(\tau) = - \Delta\epsilon_2 \left[ 1  - \theta (\tau - \tau_s) \right], \label{etasharp}
\end{equation}
where $\Delta\epsilon_2 = 6$. At time $\tau < \tau_s$, the inflaton is in the USR period with the Bunch-Davies initial vacuum. In this period, the curvature perturbation and its time derivative are given by
\begin{align}
\zeta_k(\tau) & = \frac{i H}{2 \mpl \sqrt{k^3 \epsilon_1(\tau) }} e^{-ik\tau} (1+ik\tau) \nonumber\\
& = \frac{i H}{2 \mpl \sqrt{k^3 \epsilon_1(\tau_s) }} \left( \frac{\tau_s}{\tau} \right)^3  e^{-ik\tau} (1+ik\tau) ,
\end{align}
\begin{equation}
\zeta_k'(\tau) = \frac{i H}{2 \mpl \sqrt{k^3 \epsilon_1(\tau_s) }} \left( \frac{\tau_s}{\tau} \right)^3 e^{-ik\tau} \left( k^2\tau -\frac{3}{\tau} -3 i k \right) .
\end{equation}
At time $\tau \geq \tau_s$, the inflaton is in the SR period. In this period, the curvature perturbation and its time derivative are given by 
\begin{equation}
\zeta_k(\tau) = \frac{i H}{2 \mpl \sqrt{k^3 \epsilon_1(\tau_s)}} \left[ \mathcal{A}_k e^{-ik\tau} (1+ik\tau) - \mathcal{B}_k e^{ik\tau} (1-ik\tau) \right],
\end{equation}
\begin{equation}
\zeta_k'(\tau) = \frac{i H}{2 \mpl \sqrt{k^3 \epsilon_1(\tau_s)}}  k^2 \tau \left( \mathcal{A}_k e^{-ik\tau}  - \mathcal{B}_k e^{ik\tau}  \right),
\end{equation}
where $\mathcal{A}_k$ and $\mathcal{B}_k$ are general coefficients to be determined from boundary conditions \cite{Starobinsky:1992ts, Leach:2001zf, Byrnes:2018txb, Liu:2020oqe, Tasinato:2020vdk, Karam:2022nym, Pi:2022zxs, Domenech:2023dxx}. This time, we have to consider both positive and negative frequency solutions because the boundary condition at $\tau = \tau_s$ does not correspond to Bunch-Davies vacuum. Solutions of the coefficients by requiring continuity of $\zeta_k(\tau)$ and $\zeta_k'(\tau)$ at transition $\tau = \tau_s$ are
\begin{gather}
\mathcal{A}_k = 1 + \frac{3(1 + k^2 \tau_s^2)}{2i k^3 \tau_s^3} , \\
\mathcal{B}_k = \frac{3(1 + i k \tau_s)^2}{2i k^3 \tau_s^3} e^{-2ik \tau_s}. 
\end{gather}
At the end of inflation, the power spectrum becomes
\begin{equation}
\Delta^2_{s}(k, \tau_0) = \frac{k^3}{2 \pi^2} \abs{\zeta_k(\tau_0)}^2 =  \frac{H^2}{8 \pi^2 M_{\mathrm{pl}}^2 \epsilon_1(\tau_s)} \abs{\mathcal{A}_k - \mathcal{B}_k}^2. 
\end{equation}
For a very long-wavelength modes, $\abs{k \tau_s} \ll 1$, the power spectrum approaches
\begin{equation}
\Delta^2_{s}\left( k \ll \frac{1}{\abs{\tau_s}} , \tau_0 \right) = 4 \times \frac{H^2}{8 \pi^2 M_{\mathrm{pl}}^2 \epsilon_1(\tau_s)} . \label{psusrsr}
\end{equation}

\section{Three-point functions \label{sec3}}
So far, we have considered the second-order action of the curvature perturbation that leads to two-point functions or the power spectrum at the tree level in the language of quantum field theory. Within the cosmological perturbation theory framework, higher-order correlation functions resulting from higher-order expansion of the action are expected to be small. However, because we are interested in PBH formation, where the curvature perturbation is large on certain small scales, higher-order correlation functions can be quite large. 

In this section, we discuss the third-order action that yields three-point functions of the curvature perturbation. 
Expanding \eqref{action} to third-order of $\zeta$ yields the interaction action \cite{Maldacena:2002vr}
\begin{equation}
S^{(3)}[\zeta] = S_{\mathrm{bulk}}[\zeta] + S_\mathrm{B}[\zeta] + \mpl^2  \int \md t ~\md^3 x ~2 f(\zeta) \left( \frac{\delta L}{\delta \zeta} \right)_1. \label{S3}
\end{equation}
The bulk interaction $S_{\mathrm{bulk}}[\zeta]$ reads 
\begin{align}
S_{\mathrm{bulk}}[\zeta] = \mpl^2  \int \md t ~\md^3 x ~a^3 & \left[ \epsilon_1^2  \dot{\zeta}^2 \zeta + \frac{1}{a^2} \epsilon_1^2 (\partial_i \zeta)^2 \zeta - 2 \epsilon_1  \dot{\zeta} \partial_i \zeta \partial_i \chi \right. \nonumber\\
& \left. -\frac{1}{2} \epsilon_1^3 \dot{\zeta}^2 \zeta + \frac{1}{2} \epsilon_1  \zeta (\partial_i \partial_j \chi)^2 + \frac{1}{2} \epsilon_1 \dot{\epsilon}_2 \dot{\zeta} \zeta^2 \right], \label{s3bulk}
\end{align}
where $\chi = \epsilon_1 \partial^{-2} \dot{\zeta}$. 
$S_\mathrm{B}[\zeta]$ is the surface action given by the total derivatives \cite{Arroja:2011yj, Burrage:2011hd}. The last term is interaction proportional to the equation of motion in the lowest order,
\begin{equation}
\left( \frac{\delta L}{\delta \zeta} \right)_1 = \frac{\md }{\md t} (\epsilon_1 a^3 \dot{\zeta}) - \epsilon_1 a \partial^2 \zeta.
\end{equation}
The function $f(\zeta)$ is explicitly given by
\begin{equation}
f(\zeta) = \frac{\epsilon_2}{4} \zeta^2 + \frac{\dot{\zeta}}{H} \zeta + \frac{1}{4a^2 H^2} [-(\partial_i \zeta)^2 + \partial^{-2}\partial_i \partial_j (\partial_i \zeta \partial_j \zeta)] + \frac{1}{2H}[\partial_i \zeta \partial_i \chi - \partial^{-2} \partial_i \partial_j (\partial_i \zeta \partial_j \chi)]. \label{redef}
\end{equation}

Performing field redefinition $\zeta = \bz + f(\bz)$ generates third-order terms from the second-order action \eqref{S2} that cancel relevant surface terms containing $\dot\zeta$.
As a result, we find a useful relation
\cite{Maldacena:2002vr, Arroja:2011yj, Burrage:2011hd}
\begin{equation}
S^{(2)}[\zeta] + S^{(3)}[\zeta] = S^{(2)}[\bz] + S_{\mathrm{bulk}}[\bz]. \label{sbz}
\end{equation}
In terms of $\bz$, the total action is simply given by the second-order action \eqref{S2} and bulk interaction \eqref{s3bulk}.

Three-point functions of $\zeta$ can be written schematically as
\begin{equation}
\langle \zeta_{\bk_1}(\tau) \zeta_{\bk_2}(\tau) \zeta_{\bk_3}(\tau) \rangle = \langle \bz_{\bk_1}(\tau) \bz_{\bk_2}(\tau) \bz_{\bk_3}(\tau) \rangle + \mathrm{redefinition ~terms},
\end{equation}
where the first term is generated by bulk interaction \eqref{s3bulk} and the second term is boundary contribution at $\tau$ from field redefinition.
Higher-order correction to the expectation value of an operator $\mathcal{O}(\tau)$ is calculated by the in-in perturbation theory
\begin{equation}
\langle \mathcal{O(\tau)} \rangle =  \left\langle \left[ \bar{\mathrm{T}} \exp \left( i \int_{-\infty}^{\tau} \mathrm{d}\tau' H_{\mathrm{int}}(\tau') \right) \right] \mathcal{\hat{O}} (\tau) \left[ \mathrm{T} \exp \left( -i \int_{-\infty}^{\tau} \mathrm{d\tau'} H_{\mathrm{int}}(\tau') \right) \right] \right\rangle, \label{inin}
\end{equation}
where $\mathrm{T}$ and $\bar{\mathrm{T}}$ denote time and antitime ordering. In this case, the interaction Hamiltonian is simply $H_\mathrm{int} = - \int \md^3 x \mathcal{L}_\mathrm{bulk}$, where $\mathcal{L}_\mathrm{bulk}$ is the integrand of $S_{\mathrm{bulk}}$. For three-point functions, the operator is $\bz_{\bk_1}(\tau) \bz_{\bk_2}(\tau) \bz_{\bk_3}(\tau)$. First-order expansion of \eqref{inin} reads
\begin{equation}
\langle \mathcal{O(\tau)} \rangle = 2 \int_{-\infty}^{\tau} \md\tau_1 ~\mathrm{Im} \langle \mathcal{\hat{O}} (\tau) H_{\mathrm{int}}(\tau_1) \rangle. \label{inin1}
\end{equation}
Bispectrum $\langle \! \langle \zeta_{\bk_1}(\tau) \zeta_{\bk_2}(\tau) \zeta_{\bk_3}(\tau) \rangle \! \rangle$ is defined as
\begin{equation}
\langle \zeta_{\bk_1}(\tau) \zeta_{\bk_2}(\tau) \zeta_{\bk_3}(\tau) \rangle = (2\pi)^3 \delta(\bk_1 + \bk_2 + \bk_3) \langle \! \langle \zeta_{\bk_1}(\tau) \zeta_{\bk_2}(\tau) \zeta_{\bk_3}(\tau) \rangle \! \rangle.
\end{equation}
A bispectrum has local type if it can be expressed as
\begin{align}
\langle \! \langle \zeta_{\bk_1}(\tau_0) \zeta_{\bk_2}(\tau_0) \zeta_{\bk_3}(\tau_0) \rangle \! \rangle = \frac{6}{5} f_\mathrm{NL} & \left[ \abs{\zeta_{k_1}(\tau_0)}^2 \abs{\zeta_{k_2}(\tau_0)}^2 + \abs{\zeta_{k_2}(\tau_0)}^2 \abs{\zeta_{k_3}(\tau_0)}^2 \right. \nonumber\\
& \left. + \abs{\zeta_{k_1}(\tau_0)}^2 \abs{\zeta_{k_3}(\tau_0)}^2 \right], 
\end{align}
where $f_\mathrm{NL}$ is a coefficient called nonlinear parameter.

\subsection{Slow-roll approximation \label{bispectrumsr}}
In the SR approximation, the first line in cubic self-interaction \eqref{s3bulk} has $\mathcal{O}(\epsilon^2)$ coefficients, while the second line has $\mathcal{O}(\epsilon^3)$ coefficients. At the end of inflation, when the relevant modes are in the superhorizon limit, the leading term in the field redefinition \eqref{redef} is the first term, because the other terms decay outside the horizon. Therefore, the third-order action becomes
\begin{equation}
S_{\mathrm{bulk}}[\bz] = \mpl^2  \int \md t ~\md^3 x ~a^3 \left[ \epsilon_1^2  \dot{\bz}^2 \bz + \frac{1}{a^2} \epsilon_1^2 (\partial_i \bz)^2 \zeta - 2 \epsilon_1^2  \dot{\bz} \partial_i \bz \partial_i \partial^{-2}\dot{\bz} + \mathcal{O}(\epsilon^3) \right], \label{s3bulksr}
\end{equation}
with $\zeta = \bz + \epsilon_2 \bz^2 / 4$. With the in-in perturbation theory \eqref{inin1}, we obtain bispectrum of $\bz$ as
\begin{align}
\langle \! \langle \bz_{\bk_1}(\tau_0) \bz_{\bk_2}(\tau_0) \bz_{\bk_3}(\tau_0) \rangle \! \rangle_{\mathcal{O}(\epsilon^2)} = & ~\frac{H^4}{32 \mpl^4 \epsilon_1^2 (k_1 k_2 k_3)^3} \\ 
& \times \left[ -\epsilon_{1 \star} \sum_i k_i^3 + \epsilon_{1 \star} \sum_{i \neq j} k_i k_j^2 + 8 \epsilon_{1 \star} \frac{\sum_{i > j} k_i^2 k_j^2}{\sum_i k_i} \right]. \nonumber
\end{align}
To arrive at bispectrum of $\zeta$, we have to include contribution from the field redefinition
\begin{gather}
\langle \! \langle \zeta_{\bk_1}(\tau_0) \zeta_{\bk_2}(\tau_0) \zeta_{\bk_3}(\tau_0) \rangle \!\rangle = \langle \! \langle \bz_{\bk_1}(\tau_0) \bz_{\bk_2}(\tau_0) \bz_{\bk_3}(\tau_0) \rangle \! \rangle \nonumber\\
+ \frac{1}{4} \epsilon_2(\tau_0) \times 2\left[ \abs{\zeta_{k_1}(\tau_0)}^2 \abs{\zeta_{k_2}(\tau_0)}^2 + \abs{\zeta_{k_2}(\tau_0)}^2 \abs{\zeta_{k_3}(\tau_0)}^2 + \abs{\zeta_{k_1}(\tau_0)}^2 \abs{\zeta_{k_3}(\tau_0)}^2 \right]. \label{zetabz}
\end{gather}
However, assuming $\epsilon_3$ as a constant, $\epsilon_2(\tau_0)$ diverges as $\tau_0 \rightarrow 0$. The explicit time dependence is given by
\begin{equation}
\epsilon_2(\tau_0) = \epsilon_{2 \star} \left( \frac{\tau_0}{\tau_\star} \right)^{-\epsilon_{3 \star}} \simeq \epsilon_{2 \star} \left( 1 - \epsilon_{3 \star} \log \frac{\tau_0}{\tau_\star} \right). \label{eps2eps3}
\end{equation}
At leading order of $\epsilon_2(\tau_0)$, the bispectrum of $\zeta$ is
\begin{align}
\langle \! \langle \zeta_{\bk_1}(\tau_0) \zeta_{\bk_2}(\tau_0) \zeta_{\bk_3}(\tau_0) \rangle \! \rangle_{\mathcal{O}(\epsilon^2)} = & ~\frac{H^4}{32 \mpl^4 \epsilon_1^2 (k_1 k_2 k_3)^3} \\ 
& \times \left[ (\epsilon_2 - \epsilon_1)_\star \sum_i k_i^3 + \epsilon_{ 1 \star} \sum_{i \neq j} k_i k_j^2 + 8 \epsilon_{1 \star} \frac{\sum_{i > j} k_i^2 k_j^2}{\sum_i k_i} \right]. \nonumber
\end{align}

In the squeezed limit, where one of the wavenumbers, $k_1$, is much smaller than the others, $k_1 \rightarrow 0$, the bispectrum becomes
\begin{equation}
\langle \! \langle \zeta_{\bk_1}(\tau_0) \zeta_{\bk_2}(\tau_0) \zeta_{- \bk_2}(\tau_0) \rangle \! \rangle = (2 \epsilon_1 + 4 \epsilon_2 )_\star \frac{H^4}{32 \mpl^4 \epsilon_1^2 (k_1 k_2)^3} .
\end{equation}
In terms of the mode function and spectral tilt, the squeezed bispectrum can be written as
\begin{equation}
\langle \! \langle \zeta_{\bk_1}(\tau_0) \zeta_{\bk_2}(\tau_0) \zeta_{- \bk_2}(\tau_0) \rangle \! \rangle = - (n_s (k_2, \tau_0) - 1) \abs{\zeta_{k_1}(\tau_0)}^2 \abs{\zeta_{k_2}(\tau_0)}^2 , \label{maldacena}
\end{equation}
which is called Maldacena's theorem. Here, we have shown Maldacena's theorem by explicit derivation. It can also be proven by a model-independent method by considering the evolution of the small-scale perturbation in a rescaled dS background due to the long-wavelength perturbation \cite{Creminelli:2004yq}.

Another important point discussed in \cite{Maldacena:2002vr} is the bispectrum induced by the last term in \eqref{s3bulk}
\begin{equation}
S_{\mathrm{bulk}}[\bz] \supset \frac{1}{2} \mpl^2  \int \md t ~\md^3 x ~a^3 \epsilon_1 \dot{\epsilon}_2 \dot{\bz} \bz^2 . 
\end{equation}
Although it is SR-suppressed compared to \eqref{s3bulksr}, such term induces bispectrum that evolves outside the horizon.
\begin{equation}
\langle \! \langle \bz_{\bk_1}(\tau_0) \bz_{\bk_2}(\tau_0) \bz_{\bk_3}(\tau_0) \rangle \! \rangle_{\mathcal{O}(\epsilon^3)} = \left( \epsilon_{2 \star} \epsilon_{3 \star} \log \frac{\tau_0}{\tau_\star} \right) \frac{H^4}{32 \mpl^4 \epsilon_1^2 (k_1 k_2 k_3)^3}  (k_1^3 + k_2^3 + k_3^3) ,
\end{equation}
where $\star$ denotes horizon crossing of the total wavenumber $k_1 + k_2 + k_3$. It shows that bispectrum of $\bz$ evolves outside the horizon due to $\log \tau_0$ dependence. However, for bispectrum of $\zeta$, there is an additional contribution from the field redefinition \eqref{zetabz}. After summing with the second term in \eqref{eps2eps3}, the $\log \tau_0$ dependence cancels out. Therefore, bispectrum of $\zeta$ does not evolve outside the horizon.

\subsection{Ultraslow-roll inflation}
During USR period, $\epsilon_1$ becomes extremely small because it scales as $a^{-6}$. Also, $\dot{\epsilon}_2 = 0$ because $\epsilon_2$ is a constant. Therefore, the contribution of the bulk interaction to the bispectrum is negligible and the leading contribution comes from the field redefinition. The relevant terms in the field redefinition are \cite{Namjoo:2012aa}
\begin{equation}
f(\bz) = \frac{\epsilon_2}{4} \bz^2 + \frac{\dot{\bz}}{H} \bz, \label{redefusr}
\end{equation}
which contribute
\begin{align}
\langle \! \langle \zeta_{\bk_1}(\tau_0) \zeta_{\bk_2}(\tau_0) \zeta_{\bk_3}(\tau_0) \rangle \! \rangle = 2\left( \frac{\epsilon_2}{4} + 3 \right)  & \left[ \abs{\zeta_{k_1}(\tau_0)}^2 \abs{\zeta_{k_2}(\tau_0)}^2 + \abs{\zeta_{k_2}(\tau_0)}^2 \abs{\zeta_{k_3}(\tau_0)}^2 \right. \nonumber\\
& \left. + \abs{\zeta_{k_1}(\tau_0)}^2 \abs{\zeta_{k_3}(\tau_0)}^2 \right] . 
\end{align}
Factor $3$ comes from $\dot{\zeta}_k(\tau_0) = 3 H \zeta_k(\tau_0)$, which can be derived from the USR mode function \eqref{modeusr}. Hence, we can read the local non-Gaussianity as
\begin{equation}
\frac{6}{5} f_\mathrm{NL} = 3 \Longrightarrow f_\mathrm{NL} = \frac{5}{2}.
\end{equation}
In squeezed limit, the bispectrum becomes
\begin{equation}
\langle \! \langle \zeta_{\bk_1}(\tau_0) \zeta_{\bk_2}(\tau_0) \zeta_{-\bk_2}(\tau_0) \rangle \! \rangle = 6 \abs{\zeta_{k_1}(\tau_0)}^2 \abs{\zeta_{k_2}(\tau_0)}^2, \label{squsr}
\end{equation}
which clearly violates Maldacena's theorem \eqref{maldacena}. Indeed, for a shift-symmetric action of the inflaton, Maldacena's theorem is generalized in \cite{Bravo:2017wyw, Finelli:2017fml}. The squeezed bispectrum \eqref{squsr} can be reproduced from a generalized Maldacena's theorem after substituting the USR condition.

\subsection{Ultraslow-roll to slow-roll transition}
In case there is a transition of the second SR parameter, $\dot{\epsilon_2}$ becomes large during the transition. Therefore, the leading term in the third-order action is \cite{Namjoo:2012aa, Cai:2016ngx, Chen:2013eea, Cai:2018dkf, Davies:2021loj} 
\begin{equation}
S_{\mathrm{bulk}}[\bz] = \mpl^2  \int \md \tau ~\md^3 x ~a^2 \left[ \frac{1}{2} \epsilon_1 \epsilon_2' \bz' \bz^2 + \mathcal{O}(\epsilon^2) \right]. \label{s3bulktr}
\end{equation}
Because we are interested in transition from USR to SR period, relation between $\zeta$ and $\bz$ at the end of inflation is $\zeta = \bz + \epsilon_2 \bz^2/4$. However, because $\abs{\epsilon_2(\tau_0)} \ll 1$, contribution from field redefinition to the bispectrum is much smaller than the bulk interaction \eqref{s3bulktr}. Thus, the bispectrum is given by
\begin{align}
\langle\!\langle \zeta_{\bk_1}(\tau_0) \zeta_{\bk_2}(\tau_0) \zeta_{\bk_3}(\tau_0) \rangle\!\rangle = - 2\mpl^2 & \int_{-\infty}^{\tau_0} \md \tau ~\epsilon_1(\tau) \epsilon_2'(\tau) a^2(\tau) \label{bitr} \\
& \times \mathrm{Im} \left[ \zeta_{k_1}(\tau_0) \zeta_{k_2}(\tau_0) \zeta_{k_3}(\tau_0) \zeta_{k_1}^*(\tau) \zeta_{k_2}^*(\tau) \zeta_{k_3}'^*(\tau) + \mathrm{perm} \right]. \nonumber 
\end{align}
In the next part, we will evaluate the time integral for sharp and smooth transition of $\epsilon_2(\tau)$. Similarly as in Sec.~\ref{usrsr}, we follow the parametrization of $\epsilon_2(\tau)$ in \cite{Cai:2018dkf}.

\subsubsection{Sharp transition}
Consider evolution of the second SR parameter as step function, given by \eqref{etasharp}, its time derivative is Dirac-delta function
\begin{equation}
\epsilon_2'(\tau) = \Delta\epsilon_2 \delta(\tau-\tau_s).
\end{equation}
Substituting it to the time integral in \eqref{bitr} leads to
\begin{align}
\langle\!\langle \zeta_{\bk_1}(\tau_0) \zeta_{\bk_2}(\tau_0) \zeta_{\bk_3}(\tau_0) \rangle\!\rangle = & - 2\mpl^2 \epsilon_1(\tau_s) a^2(\tau_s)  \Delta\epsilon_2 \\
& \times \mathrm{Im} \left[ \zeta_{k_1}(\tau_0) \zeta_{k_2}(\tau_0) \zeta_{k_3}(\tau_0) \zeta_{k_1}^*(\tau_s) \zeta_{k_2}^*(\tau_s) \zeta_{k_3}'^*(\tau_s) + \mathrm{perm} \right]. \nonumber
\end{align}
Then, we can substitute $\zeta_k(\tau)$ and $\zeta_k'(\tau)$ given in Sec.~\ref{s2sharp} to obtain the bispectrum. Although the general form of bispectrum is complicated, in the limit $k_1, k_2, k_3 \ll k_s$ the bispectrum becomes
\begin{align}
\langle\!\langle \zeta_{\bk_1}(\tau_0) \zeta_{\bk_2}(\tau_0) \zeta_{\bk_3}(\tau_0) \rangle\!\rangle = & ~2\mpl^2 \epsilon_1(\tau_s) a^2(\tau_s) \Delta\epsilon_2 ~ \left( \frac{H^2}{4 \mpl^2 \epsilon_1(\tau_s)} \right)^3 \nonumber \\
& \times \frac{4}{k_s^2} \left( \frac{1}{k_1^3 k_2^3} + \frac{1}{k_2^3 k_3^3} + \frac{1}{k_1^3 k_3^3} \right), 
\end{align}
which is a local type bispectrum. Thus, we can read the nonlinear parameter $ f_\mathrm{NL} = 5/8$.

\subsubsection{Smooth transition}
Consider evolution of the second SR parameter given by \eqref{etasmooth}, its time derivative  is
\begin{equation}
\epsilon_2'(\tau) = \frac{-2 s^2(s-3)(s+3)  }{\tau \left[ \left( \frac{\tau_s}{\tau} \right)^s (s-3) + s + 3 \right]^2} \left( \frac{\tau_s}{\tau} \right)^s. \label{etapsmooth}
\end{equation}
Because  the second SR parameter is  constant for $\tau < \tau_s$, the domain of time integral in \eqref{bitr} is simplified to 
\begin{align}
\langle\!\langle \zeta_{\bk_1}(\tau_0) \zeta_{\bk_2}(\tau_0) \zeta_{\bk_3}(\tau_0) \rangle\!\rangle = - 2\mpl^2 & \int_{\tau_s}^{\tau_0} \md \tau ~\epsilon_1(\tau) \epsilon_2'(\tau) a^2(\tau) \\
& \times \mathrm{Im} \left[ \zeta_{k_1}(\tau_0) \zeta_{k_2}(\tau_0) \zeta_{k_3}(\tau_0) \zeta_{k_1}^*(\tau) \zeta_{k_2}^*(\tau) \zeta_{k_3}'^*(\tau) + \mathrm{perm} \right]. \nonumber
\end{align}
Unfortunately, the general form of bispectrum is complicated. However, it becomes simple in the squeezed limit $k_1 \ll k_2 = k_3$ for perturbation modes that cross the horizon much before the transition $k_1, k_2, k_3 \ll k_s$. Substituting \eqref{epssmooth}, \eqref{etapsmooth}, and \eqref{zetasmooth} to the time integral leads to
\begin{align}
\langle\!\langle \zeta_{\bk_1}(\tau_0) \zeta_{\bk_2}(\tau_0) \zeta_{-\bk_2}(\tau_0) \rangle\!\rangle & =  \abs{\zeta_{k_1}(\tau_0)}^2 \abs{\zeta_{k_2}(\tau_0)}^2 \int_{\tau_s}^{\tau_0} \md\tau \sqrt{\frac{\epsilon_1(\tau_0)}{\epsilon_1(\tau)}} \epsilon_2'(\tau) \left( 1 + \frac{1}{2} \epsilon_2(\tau) \right) \nonumber\\
& = -2 \eta_V \sqrt{\frac{\epsilon_1(\tau_0)}{\epsilon_1(\tau_s)}}  \abs{\zeta_{k_1}(\tau_0)}^2 \abs{\zeta_{k_2}(\tau_0)}^2.
\end{align}
Thus, we can read off the nonlinear parameter
\begin{equation}
\frac{12}{5} f_\mathrm{NL} = -2 \eta_V \sqrt{\frac{\epsilon_1(\tau_0)}{\epsilon_1(\tau_s)}} ,
\end{equation}
which is suppressed by SR parameter $\eta_V$. This means that the bispectrum generated by the transition \eqref{etasmooth} is negligible. In \cite{Kristiano:2024vst}, it is argued that suppression in the bispectrum is a result of \eqref{etasmooth} satisfying Wands duality condition \cite{Wands:1998yp}, which makes the integrand of \eqref{s3bulktr} a total time derivative. The squeezed bispectrum in both smooth and sharp transition cases can be reproduced from the generalized Maldacena's theorem by considering only background evolution \cite{Namjoo:2023rhq} following \cite{Creminelli:2004yq}.

\section{Generation of large fluctuation \label{sec4}}
In Sec.~\ref{usr}, we have shown that violation of the SR approximation, especially the USR condition, leads to the growth of curvature perturbation outside the horizon \cite{Yokoyama:1998rw,Saito:2008em}. Although the USR period cannot realize the required e-folds number of inflation, it can be useful for other purposes. If the inflaton undergoes USR period only for a short duration, the curvature perturbation will be enhanced only within that duration. At the end of inflation, it corresponds to amplification of the curvature perturbation at a range of wavenumbers that cross the horizon during the USR period.

Assuming sharp transitions, evolution of the second SR parameter corresponding to the potential in Fig.~\ref{fig1} is shown by the brown curve in Fig.~\ref{fig2}. The inflaton undergoes USR period in the flat region of potential from time $t_s$ to $t_e$. In order to explain the observed CMB fluctuations, at early time, $t \lesssim t_s$, the inflation was in SR period with slightly tilted potential. At late time, $t \gtrsim t_e$, there is no observational constraint on the potential. A potential is viable as long as the curvature perturbation does not grow during the final period.

In this section, we study the power spectrum and bispectrum of curvature perturbation in such a setup.
\subsection{Power spectrum}
At early time, $t \lesssim t_s$, the inflaton was in SR period  with Bunch-Davies initial vacuum, so the curvature perturbation is given by
\begin{equation}
\zeta_k(\tau) = \left( \frac{i H}{2 \mpl \sqrt{k^3 \epsilon_{1s}}} \right)_\star  e^{-ik\tau} (1+ik\tau), \label{zetasr}
\end{equation}
where $\epsilon_{1 s}$ is the first SR parameter during this SR period.

At $t_s \lesssim t \lesssim t_e$, the inflaton is in the intermediate USR period. We define $\tau_s$ and $\tau_e$ as the conformal time corresponding to $t_s$ and $t_e$, respectively. The first SR parameter can be written as $\epsilon_1(\tau) = \epsilon_{1 s} (\tau / \tau_s)^6$ based on proportionality in \eqref{epsusr}. Therefore, the curvature perturbation becomes
\begin{equation}
\zeta_k(\tau) = \left( \frac{i H}{2 \mpl \sqrt{k^3 \epsilon_{1 s}}} \right)_\star  \left( \frac{\tau_s}{\tau} \right)^3  \left[ \mathcal{A}_k e^{-ik\tau} (1+ik\tau) - \mathcal{B}_k e^{ik\tau} (1-ik\tau) \right], \label{zetausr}
\end{equation}
where coefficients $\mathcal{A}_k$ and $\mathcal{B}_k$ are determined by matching to the SR solution \eqref{zetasr} at the boundary. Solutions of the coefficients by requiring continuity of $\zeta_k(\tau)$ and $\zeta_k'(\tau)$ at transition $\tau = \tau_s$ are
\begin{gather}
\mathcal{A}_k = 1 - \frac{3(1 + k^2 \tau_s^2)}{2i k^3 \tau_s^3} \label{coefa2}, \\
\mathcal{B}_k = - \frac{3(1 + i k \tau_s)^2}{2i k^3 \tau_s^3} e^{-2ik \tau_s}. \label{coefb2}
\end{gather}

At late time, $t \gtrsim t_e$, the inflaton goes back to SR dynamics. The curvature perturbation can be written as
\begin{equation}
\zeta_k(\tau) = \left( \frac{i H}{2 \mpl \sqrt{k^3 \epsilon_{1 s}}} \right)_\star  \left( \frac{\tau_s}{\tau_e} \right)^3  \left[ \mathcal{C}_k e^{-ik\tau} (1+ik\tau) - \mathcal{D}_k e^{ik\tau} (1-ik\tau) \right], \label{zetasr2}
\end{equation}
where coefficients $\mathcal{C}_k$ and $\mathcal{D}_k$ are determined by matching to the USR solution \eqref{zetausr} at the boundary. Solutions of the coefficients by requiring continuity of $\zeta_k(\tau)$ and $\zeta_k'(\tau)$ at transition $\tau = \tau_e$ are
\begin{align}
\mathcal{C}_k =  \frac{-1}{4k^6 \tau_s^3 \tau_e^3} & \left\lbrace 9(k\tau_s - i)^2 (k\tau_e + i)^2 e^{2ik(\tau_e - \tau_s)} \right. \nonumber\\
& \left. - \left[k^2 \tau_s^2 (2 k \tau_s + 3i) + 3i \right] \left[k^2 \tau_e^2 (2k \tau_e - 3i) - 3i \right] \right\rbrace, \label{coefa3}
\end{align}
\begin{align}
\mathcal{D}_k = \frac{3}{4k^6 \tau_s^3 \tau_e^3} & \left\lbrace e^{-2i k \tau_s} [3 + k^2 \tau_e^2 (3-2i k \tau_e)] (k \tau_s - i)^2 \right. \nonumber\\
& \left. + i e^{-2i k \tau_e} \left[ 3i + k^2 \tau_s^2 (2 k \tau_s + 3i) \right] (k \tau_e - i)^2 \right\rbrace. \label{coefb3}
\end{align}
 As before, $k_s$ and $k_e$ are wavenumbers which cross the horizon at $\tau_s$ and $\tau_e$, respectively. At the end of inflation, $\tau_0~(\rightarrow 0)$, the tree-level power spectrum is
\begin{equation}
\Delta^2_{s(0)}(k, \tau_0) = \frac{k^3}{2 \pi^2} \abs{\zeta_k(\tau_0)}^2 = \left( \frac{H^2}{8 \pi^2 M_{\mathrm{pl}}^2 \epsilon_{1 s}} \right)_\star \left( \frac{k_e}{k_s} \right)^6 \abs{\mathcal{C}_k - \mathcal{D}_k}^2. \label{ps0}
\end{equation}

On large scale, the power spectrum approaches an almost scale-invariant limit
\begin{equation}
\Delta_{s(\mathrm{SR})}^{2}(k) \equiv \Delta^2_{s(0)}(k \ll k_s, \tau_0) = \left( \frac{H^2}{8 \pi^2 M_{\mathrm{pl}}^2 \epsilon_{1 s}} \right)_\star \left[ 1 + \mathcal{O}\left( \frac{k^2}{k_s^2} \right) \right],
\end{equation}
which is consistent with CMB observation. Assuming $k_e \gg k_s$, the next order approximation is
\begin{equation}
\Delta^2_{s(0)}(k, \tau_0) \simeq \left( \frac{H^2}{8 \pi^2 M_{\mathrm{pl}}^2 \epsilon_{1 s}} \right)_\star \left[ 1 - \frac{8}{5} \left( \frac{k_e}{k_s} \right)^3 \left( \frac{k}{k_s} \right)^2 + \frac{24}{35} \left( \frac{k_e}{k_s} \right)^3 \left( \frac{k}{k_s} \right)^4 + \mathcal{O}\left( \frac{k^6}{k_s^6} \right) \right].
\end{equation}
The minus sign in the coefficient of $(k/k_s)^2$ is important because it can explain a dip in the power spectrum before amplification start. As $k$ approaches $k_s$, the $k^2$-term grows until it reaches unity. At this point, the $k^4$-term is not large enough yet, so the power spectrum almost vanishes. This yields a dip in the power spectrum, which happens at the wavenumber
\begin{equation}
    k_\mathrm{dip} \simeq k_s \left[ \frac{5}{8} \left( \frac{k_s}{k_e} \right)^3 \right]^{1/2}. \label{kdip}
\end{equation}
It has been argued that 
this dip may lead to an interesting observational consequence \cite{Balaji:2022zur}.

At $k_\mathrm{dip} < k < k_s$, the $k^4$-term grows while the $k^6$-term is still small, which explains the apparent $k^4$-growth in the power spectrum \cite{Byrnes:2018txb}. On a small scale with a larger wavenumber, $k_s \lesssim k \lesssim k_e$, the power spectrum is oscillating around
\begin{equation}
4 \times \Delta_{s(\mathrm{PBH})}^{2} \equiv 4 \times \Delta_{s(\mathrm{SR})}^{2}(k_s) \left( \frac{k_e}{k_s} \right)^6,
\end{equation}
whose high-density peak may collapse into PBHs. Numerical coefficient 4 appears for the same reason as \eqref{psusrsr}. It is amplified by a factor of $(k_e/k_s)^6$ compared to the CMB-scale power spectrum. To generate a significant abundance of PBHs, typically peak with $\mathcal{O}(0.01)$ on small-scale power spectrum is required. It means that the power spectrum is amplified $\mathcal{O}(10^7)$ compared to CMB-scale. In terms of the number of e-folds during the USR period $N_\mathrm{USR} = \log[ a(\tau_e)/ a(\tau_s) ]$, we need to have 
\begin{equation}
N_\mathrm{USR} = \log \frac{\tau_s}{\tau_e} = \log \frac{k_e}{k_s} \simeq \frac{1}{6} \log 10^7 = 2.68,
\end{equation}
to achieve a significant abundance of PBHs.
\begin{figure}[tbp]
\centering 
\includegraphics[width=0.7\textwidth]{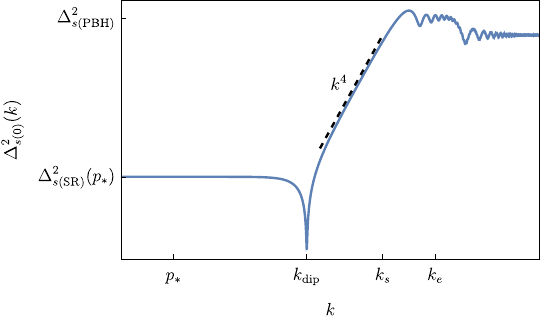}
\caption{\label{fig4} Power spectrum of the curvature perturbation. At CMB pivot scale, $p_* = 0.05 ~\mathrm{Mpc}^{-1}$, the power spectrum is almost scale invariant with amplitude $\Delta_{s(\mathrm{SR})}^{2}(p_*) = 2.1 \times 10^{-9}$, based on observational result \cite{Planck:2018jri}. At $k_\mathrm{dip} < k < k_s$, the power spectrum grows with function $k^4$ until it reaches the peak. At $k_s < k <k_e$, the power spectrum is oscillating around $4 \times \Delta_{s(\mathrm{PBH})}^{2}$. Similar behavior at $k > k_e$ because SR period is imposed after the USR period.}
\end{figure}

In this model, we can see that the power spectrum remains amplified even for $k > k_e$, which oscillates around $\Delta_{s(\mathrm{PBH})}^{2}$. This happens because the inflaton goes directly to the SR period after the USR period. In case the inflaton undergoes a constant-roll period with $\epsilon_2 > 0$ after the USR period, the power spectrum at the end of inflation will decrease for $k > k_e$.

In this review, we choose an USR period for the intermediate phase as an example. In principle, the intermediate phase can be generalized to a constant-roll period as long as the curvature perturbation grows outside the horizon. The requirement was discussed below Eq. \eqref{crgrow}. However, for a general constant-roll period, the mode function becomes the Hankel function, which is given by \eqref{mshankel} with \eqref{crnu}. Technically, it becomes more difficult to match the coefficient of constant-roll period with the SR period because it involves derivative of Hankel function. In \cite{Tasinato:2020vdk}, a method to approximate the mode function is proposed by utilizing the fact that the duration of the SR violating regime is extremely short compared to the total duration of inflation. An inflation model with two phases of SR violating period is also introduced in \cite{Tasinato:2020vdk} that yields $k^8$ growth in the power spectrum, which is much steeper than $k^4$ growth.

\subsection{Bispectrum}
Consider the evolution of the second SR parameter shown by the brown curve in Fig.~\ref{fig2}. Its time derivative can be modelled as Dirac-delta function
\begin{equation}
\epsilon_2'(\tau) = \Delta\epsilon_2 [ -\delta(\tau - \tau_s) + \delta(\tau - \tau_e) ]. 
\end{equation}
Substituting it to the time integral in \eqref{bitr} leads to
\begin{gather}
\langle\!\langle \zeta_{\bk_1}(\tau_0) \zeta_{\bk_2}(\tau_0) \zeta_{\bk_3}(\tau_0) \rangle\!\rangle = 2\mpl^2 \epsilon_1(\tau_s) a^2(\tau_s) \Delta\epsilon_2 \mathrm{Im} \left[ \zeta_{k_1}(\tau_0) \zeta_{k_2}(\tau_0) \zeta_{k_3}(\tau_0) \zeta_{k_1}^*(\tau_s) \zeta_{k_2}^*(\tau_s) \zeta_{k_3}'^*(\tau_s) \right] \nonumber\\
- 2\mpl^2 \epsilon_1(\tau_e) a^2(\tau_e) \Delta\epsilon_2 \mathrm{Im} \left[ \zeta_{k_1}(\tau_0) \zeta_{k_2}(\tau_0) \zeta_{k_3}(\tau_0) \zeta_{k_1}^*(\tau_e) \zeta_{k_2}^*(\tau_e) \zeta_{k_3}'^*(\tau_e) \right] + \mathrm{perm}. 
\end{gather}
In squeezed limit, the bispectrum becomes 
\begin{gather}
\langle \! \langle \zeta_{\bk_1}(\tau_0) \zeta_{\bk_2}(\tau_0) \zeta_{-\bk_2}(\tau_0) \rangle \! \rangle = - \left\lbrace 4 \Delta\epsilon_2 \mpl^2 \epsilon_1(\tau_e) a^2(\tau_e) \mathrm{Im} \left[ \frac{\zeta_{k_2}^2(\tau_0)}{\abs{\zeta_{k_2}(\tau_0)}^2} \zeta_{k_2}^*(\tau_e) \zeta_{k_2}'^*(\tau_e) \right] \right. \nonumber\\ 
\left. - 4 \Delta\epsilon_2 \mpl^2 \epsilon_1(\tau_s) a^2(\tau_s) \mathrm{Im} \left[ \frac{\zeta_{k_2}^2(\tau_0)}{\abs{\zeta_{k_2}(\tau_0)}^2} \zeta_{k_2}^*(\tau_s) \zeta_{k_2}'^*(\tau_s) \right] \right\rbrace \abs{\zeta_{k_1}(\tau_0)}^2 \abs{\zeta_{k_2}(\tau_0)}^2 .
\end{gather}
We define the term inside parenthesis as $C_0(k)$, so the bispectrum can be written as
\begin{equation}
\langle \! \langle \zeta_{\bk_1}(\tau_0) \zeta_{\bk_2}(\tau_0) \zeta_{-\bk_2}(\tau_0) \rangle \! \rangle = -  C_0(k_2) \abs{\zeta_{k_1}(\tau_0)}^2 \abs{\zeta_{k_2}(\tau_0)}^2 .
\end{equation}
We can show that $C_0(k) = n_s(k, \tau_0) - 1$ analytically, which confirms Maldacena's theorem explicitly \cite{Kristiano:2023scm}.
\begin{figure}[tbp]
\centering 
\includegraphics[width=0.75\textwidth]{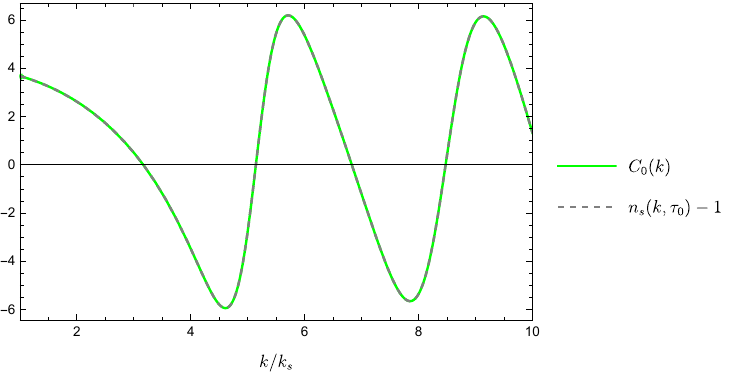}
\caption{\label{fig5} Plot of $C_0(k)$ and $n_s(k, \tau_0) - 1$. We choose $k_e/k_s = 10$ only for illustrative purposes.}
\end{figure}

\begin{figure}[tbp]
\centering 
\includegraphics[width=0.75\textwidth]{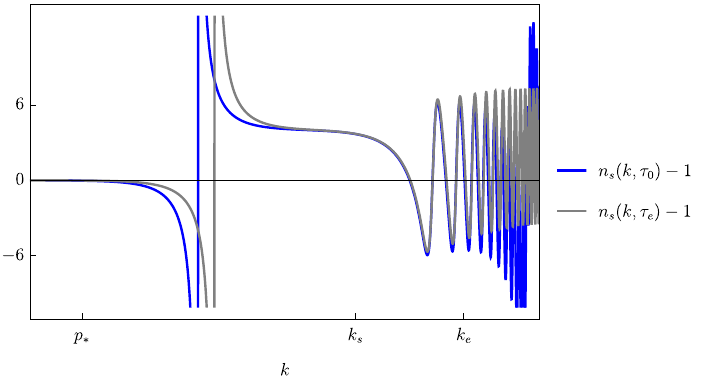}
\caption{\label{fig6} Plot of $n_s - 1$ at the end of USR period and at the end of inflation. For a range of wavenumber between $p_*$ and $k_s$, the spectral tilt evolves, altough such perturbations already far outside the horizon at the end of USR period.}
\end{figure}

At the end of inflation, we have shown that only the bulk interaction \eqref{s3bulktr} contributes to the bispectrum. During the USR period, the bispectrum also satisfies Maldacena's theorem, where both the bulk interaction \eqref{s3bulktr} and field redefinition \eqref{redefusr} are important. Therefore, in general time, the squeezed bispectrum is
\begin{equation}
\langle \! \langle \zeta_{\bk_1}(\tau) \zeta_{\bk_2}(\tau) \zeta_{-\bk_2}(\tau) \rangle \! \rangle = - ( n_s(k_2 , \tau) - 1) \abs{\zeta_{k_1}(\tau)}^2 \abs{\zeta_{k_2}(\tau)}^2 . \label{constrans}
\end{equation}
We can read that spectral tilt of the short-scale perturbation, depends on time even at superhorizon scale, specifically around the transition time. Plot of the spectral tilt at the end of USR period and at the end of inflation are shown in Fig.~\ref{fig6}. Thus, in contrast to the case of the SR inflation in Sec.~\ref{bispectrumsr} where time dependence $\log \tau_0$ is cancelled, the squeezed bispectrum evolves outside the horizon in this model. In case all wavenumbers have magnitude around the peak scale, the bispectrum is expected to deviate significantly from Maldacena's limit.

In the presence of peaks and troughs in the power spectrum, Maldacena's theorem is generalized in \cite{Namjoo:2024ufv}. In addition, the consistency relation \eqref{constrans} fixes the sign of $f_\mathrm{NL}$ at the peak, which has an implication for the abundances of PBH \cite{Firouzjahi:2023xke}.

\subsection{Beyond tree-level contribution}
So far, we have discussed two-point and three-point functions of the curvature perturbation at tree-level approximation. We have shown that the sharp transition of the second SR parameter generates a $\mathcal{O}(1)$ contribution to the nonlinear parameter $f_\mathrm{NL}$. It is a large contribution to the bispectrum, which implies that a higher-order correction than first-order perturbation theory can also be large.

The second-order expansion of in-in perturbation theory reads
\begin{gather}
\langle \mathcal{O(\tau)} \rangle_{(1)} = \langle \mathcal{O(\tau)} \rangle_{(0,2)}^\dagger + \langle \mathcal{O(\tau)} \rangle_{(1,1)} + \langle \mathcal{O(\tau)} \rangle_{(0,2)}, \label{secondinin} \\
\langle \mathcal{O(\tau)} \rangle_{(1,1)} = \int_{-\infty}^{\tau} \mathrm{d}\tau_1 \int_{-\infty}^{\tau} \mathrm{d}\tau_2 \left\langle H_{\mathrm{int}}(\tau_1) \mathcal{\hat{O}} (\tau) H_{\mathrm{int}}(\tau_2) \right\rangle, \nonumber\\
\langle \mathcal{O(\tau)} \rangle_{(0,2)} = - \int_{-\infty}^{\tau} \mathrm{d}\tau_1 \int_{-\infty}^{\tau_1} \mathrm{d}\tau_2 \left\langle \mathcal{\hat{O}} (\tau) H_{\mathrm{int}}(\tau_1) H_{\mathrm{int}}(\tau_2) \right\rangle. \nonumber
\end{gather}
In the presence of cubic self-interaction in the Hamiltonian, the second-order perturbation contributes to one-loop correction to the power spectrum and tree-level contribution to the trispectrum. For an inflation model with sharp transition of the second SR parameter, a one-loop correction to the large-scale power spectrum was first derived in \cite{Kristiano:2022maq}. In this case, the operator is $\zeta_{\bp}(\tau) \zeta_{-\bp}(\tau) $, where $\mathbf{p}$ is the CMB scale wavevector, evaluated at $\tau=\tau_0 ~(\rightarrow 0)$. 

At the end of inflation, there is no contribution from the field redefinition because it is evaluated at the SR period. Only the bulk interaction \eqref{s3bulktr} contributes to the one-loop correction. Substituting the corresponding Hamiltonian of \eqref{s3bulktr} to \eqref{secondinin} yields
\begin{align}
\langle \zeta_{\bp}(\tau_0) \zeta_{-\bp}(\tau_0) \rangle_{(1,1)} = & ~\frac{1}{4} \mpl^4 \int_{-\infty}^{\tau_0} \md \tau_1 ~a^2(\tau_1) \epsilon_1(\tau_1) \epsilon_2'(\tau_1) \int_{-\infty}^{\tau_0} \md \tau_2 ~a^2(\tau_2) \epsilon_1(\tau_2) \epsilon_2'(\tau_2)  \nonumber \\
& \times \int \prod_{a = 1}^6 \left[ \frac{\md^3 k_a}{(2\pi)^3} \right] \delta(\bk_1+\bk_2+\bk_3) \delta(\bk_4+\bk_5+\bk_6) \nonumber\\
& \times \left\langle \zeta_{\bk_1}'(\tau_1) \zeta_{\bk_2}(\tau_1) \zeta_{\bk_3}(\tau_1) \zeta_{\bp}(\tau_0) \zeta_{-\bp}(\tau_0) \zeta_{\bk_4}'(\tau_2) \zeta_{\bk_5}(\tau_2) \zeta_{\bk_6}(\tau_2) \right\rangle, \label{onel1}
\end{align}
\begin{align}
\langle \zeta_{\bp}(\tau_0) \zeta_{-\bp}(\tau_0) \rangle_{(0,2)} = & -\frac{1}{4} \mpl^4 \int_{-\infty}^{\tau_0} \md \tau_1 ~a^2(\tau_1) \epsilon_1(\tau_1) \epsilon_2'(\tau_1) \int_{-\infty}^{\tau_1} \md \tau_2 ~a^2(\tau_2) \epsilon_1(\tau_2) \epsilon_2'(\tau_2) \nonumber \\
& \times \int \prod_{a = 1}^6 \left[ \frac{\md^3 k_a}{(2\pi)^3} \right] \delta(\bk_1+\bk_2+\bk_3) \delta(\bk_4+\bk_5+\bk_6) \nonumber\\
& \times \left\langle \zeta_{\bp}(\tau_0) \zeta_{-\bp}(\tau_0) \zeta_{\bk_1}'(\tau_1) \zeta_{\bk_2}(\tau_1) \zeta_{\bk_3}(\tau_1) \zeta_{\bk_4}'(\tau_2) \zeta_{\bk_5}(\tau_2) \zeta_{\bk_6}(\tau_2) \right\rangle. \label{onel2}
\end{align}
Performing Wick contraction and summing both terms lead to
\begin{align}
& \langle \! \langle \zeta_{\bp}(\tau_0) \zeta_{-\bp}(\tau_0) \rangle \! \rangle_{(1)} = \frac{1}{4} \mpl^4 \epsilon_1^2(\tau_e) a^4(\tau_e) (\Delta\epsilon_2)^2 \int \frac{\md^3 k}{(2\pi)^3} \left\lbrace \zeta_p(\tau_0) \zeta_p^*(\tau_0) \left( 4 \zeta_p' \zeta_p'^* \zeta_q \zeta_q^* \zeta_k \zeta_k^* \right. \right. \nonumber\\ 
& \left. \left. + 8 \zeta_p'^* \zeta_p \zeta_q' \zeta_q^* \zeta_k \zeta_k^*  + 8 \zeta_p' \zeta_p^* \zeta_q'^* \zeta_q \zeta_k \zeta_k^* + 16 \zeta_p \zeta_p^* \zeta_q' \zeta_q'^* \zeta_k \zeta_k^* \right)_{\tau = \tau_e} - \mathrm{Re} \left[ \zeta_p(\tau_0)\zeta_p(\tau_0)  \right. \right. \\ 
& \left. \left. \times \left( 4  \zeta_p'^* \zeta_p'^* \zeta_q \zeta_q^* \zeta_k \zeta_k^* + 8  \zeta_p'^* \zeta_p^* \zeta_q' \zeta_q^* \zeta_k \zeta_k^* + 8  \zeta_p'^* \zeta_p^* \zeta_q'^* \zeta_q \zeta_k \zeta_k^* + 16 \zeta_p^* \zeta_p^* \zeta_q' \zeta_q'^* \zeta_k \zeta_k^* \right)_{\tau = \tau_e} \right] \right\rbrace. \nonumber
\end{align}
 Here, we only show the leading contribution to the one-loop correction that comes from integral of Dirac-delta function at $\tau = \tau_e$. The contributions of $\tau = \tau_s$ are subleading, which are explicitly evaluated in \cite{Kristiano:2023scm}.

For $p \ll k $, the one-loop correction can be simplified to
\begin{equation}
\langle \! \langle \zeta_{\bp}(\tau_0) \zeta_{-\bp}(\tau_0) \rangle \! \rangle_{(1)} = \frac{1}{4} \mpl^4 (\Delta\epsilon_2)^2 \abs{\zeta_p(\tau_0)}^2 \int \frac{\md^3 k}{(2\pi)^3} 16 \left[\epsilon_1^2  a^4\abs{\zeta_k}^2 \mathrm{Im}(\zeta_p \zeta_p'^*) \mathrm{Im}(\zeta_k \zeta_k'^*) \right]_{\tau = \tau_e} . \label{loop}
\end{equation}
We find that the one-loop correction is exactly proportional to the quantum commutator \eqref{zetacom}, which implies the importance of decaying mode even for superhorizon mode function. If we only consider the constant mode of curvature perturbation at superhorizon scale, we will obtain an incorrect result $\left[ \hat{\zeta}_{\bp} (\tau),\hat{\zeta}_{\bq}' (\tau) \right] = 0$. From commutator \eqref{zetacom}, we find
\begin{equation}
\mathrm{Im}(\zeta_{k}(\tau) \zeta_{k}'^*(\tau)) = \frac{1}{4 \mpl^2 \epsilon(\tau) a^2(\tau) }, \label{imzeta}
\end{equation} 
which does not depend on wavenumber $k$.  As we have already seen, in the standard SR inflation this quantity decreases exponentially as a function of conformal time.
However, during USR regime, since
$\epsilon$ decreases much more rapicly than $a^{-2}$, the right-hand-side of \eqref{zetacom} and \eqref{imzeta} {\it increases} exponentially, which makes the one-loop correction to the power spectrum large even on the large scale probed by CMB.

Then, substituting \eqref{imzeta} into \eqref{loop}, we obtain a one-loop correction to the power spectrum
\begin{equation}
\Delta_{s(1)}^2(p, \tau_0) = \frac{1}{4} (\Delta\epsilon_2)^2  \Delta_{s(0)}^{2}(p, \tau_0)  \int_{k_{\mathrm{IR}}}^{k_{\mathrm{UV}}} \frac{\md k}{k} \Delta_{s(0)}^{2}(k, \tau_e). \label{loop2}
\end{equation}
Because we are interested in the effect of large fluctuation on small scales related to PBH formation to the fluctuations on large scales observed by CMB, a natural choice of cutoffs are $k_{\mathrm{IR}} = k_s$ and $k_{\mathrm{UV}} = k_e$. For $k_e/k_s \ll 1$, evaluating the wavenumber integral with small-scale mode function \eqref{zetausr} yields
\begin{equation}
\Delta_{s(1)}^2(p, \tau_0) = \frac{1}{4} (\Delta\epsilon_2)^2  \Delta_{s(0)}^{2}(p, \tau_0) \Delta_{s(\mathrm{PBH})}^{2} \left( 1.1 + \log\frac{k_e}{k_s} \right).
\end{equation}
Requiring the one-loop correction to be much smaller than tree-level contribution yields an upper-bound on small-scale power spectrum
\begin{equation}
\Delta_{s(\mathrm{PBH})}^2 \ll \frac{1}{(\Delta\epsilon_2)^2} \simeq 0.03, \label{bound}
\end{equation}
where the numerical value is obtained by substituting $\Delta\epsilon_2 = 6$ for transition from USR to SR period. Such a bound suggests that sharp transition of the second SR parameter cannot generate large fluctuation on small scales, otherwise perturbation theory breaks down. This result should be understood under caveat that there are other contributions from quartic Hamiltonian, both induced by the third-order action \eqref{s3bulktr} and fourth-order action.

\section{Discussion and summary \label{sec5}}
\subsection{Discussion}
Following \cite{Karam:2022nym}, PBH formation models in single-field inflation can be classified into two categories. The first category is models with features in the inflationary potential or nonminimally coupled inflaton with potential defined in the Einstein frame. Models with an extremely flat feature or an inflection point in the potential fall in this category \cite{Ivanov:1994pa, Kawasaki:1997ju, Kawasaki:1998vx, Kawasaki:2016pql, Gao:2018pvq, Nanopoulos:2020nnh, Wu:2021zta, Stamou:2021qdk, Cheng:2018yyr, Ballesteros:2019hus, Pi:2017gih, Dalianis:2019asr, Dalianis:2018frf, Mahbub:2019uhl, Fu:2022ypp, Cicoli:2018asa, Ozsoy:2018flq, Cicoli:2022sih, Ezquiaga:2017fvi, Ballesteros:2017fsr, Drees:2019xpp, Cheong:2019vzl, Rasanen:2018fom, Garcia-Bellido:2017mdw, Ragavendra:2020sop}. These models can be derived from high-energy theories such as supergravity \cite{Kawasaki:1997ju, Kawasaki:1998vx, Kawasaki:2016pql, Gao:2018pvq, Nanopoulos:2020nnh, Wu:2021zta, Stamou:2021qdk}, axion monodromy \cite{Cheng:2018yyr, Ballesteros:2019hus}, scalaron in $R^2$-gravity \cite{Pi:2017gih}, $\alpha$-attractor \cite{Dalianis:2019asr, Dalianis:2018frf, Mahbub:2019uhl, Fu:2022ypp}, and string theory \cite{Cicoli:2018asa, Ozsoy:2018flq, Cicoli:2022sih}. Also, there are models from Higgs inflation which do not require theories beyond the standard model \cite{Ezquiaga:2017fvi, Ballesteros:2017fsr, Drees:2019xpp, Cheong:2019vzl, Rasanen:2018fom}. Other examples of features are a tiny bump or dip \cite{Mishra:2019pzq, Atal:2019cdz, Zheng:2021vda, Wang:2021kbh, Rezazadeh:2021clf, Iacconi:2021ltm}, an upward or downward step \cite{Cai:2021zsp, Kefala:2020xsx, Ng:2021hll, Motohashi:2019rhu, Inomata:2021uqj, Inomata:2021tpx, Kawaguchi:2023mgk}, polynomial shape \cite{Hertzberg:2017dkh, Ballesteros:2020qam, Kannike:2017bxn, Di:2017ndc}, and Coleman-Weinberg potential \cite{Yokoyama:1998pt, Saito:2008em, Bugaev:2008bi, Karam:2023haj}. In these examples, modification of potential makes a sharp transition of the second SR parameter on the inflaton dynamics.

The second category is models with modified gravity or beyond nonminimally coupled inflaton. Examples for this category are models based on $k$ \cite{Armendariz-Picon:1999hyi} or $G$ \cite{Kobayashi:2011nu} inflation \cite{Solbi:2021wbo, Solbi:2021rse, Teimoori:2021pte, Heydari:2023xts, Heydari:2023rmq, Heydari:2024bxj}, the effective field theory of inflation \cite{Ballesteros:2018wlw, Ballesteros:2021fsp, Ashoorioon:2019xqc}, $f(R)$ gravity \cite{Frolovsky:2022ewg}, a nonminimal derivative coupling \cite{Fu:2019ttf, Heydari:2021gea, Heydari:2021qsr}, Gauss-Bonnet inflation \cite{Kawai:2021edk, Kawaguchi:2022nku}, and bumpy axion inflation \cite{Ozsoy:2020kat}. In these examples, amplification of small-scale perturbation can be realized by a sharp transition of the second SR parameter and/or other parameters. In \cite{Ballesteros:2018wlw, Ballesteros:2021fsp}, amplification of small-scale perturbation is caused by a sharp transition of the sound speed, $c_s$, a quantity that parametrizes deviation from a canonical kinetic term. In this case, coupling $\epsilon_2'$ in cubic self-interaction \eqref{s3bulktr} is modified to $(\epsilon_2 / c_s^2)'$ \cite{Chen:2006nt, Seery:2005wm}. For this category, rapid slowdown of the inflaton's velocity might not be the only reason of amplification of the curvature perturbation.

There have been some works following \cite{Kristiano:2022maq, Kristiano:2023scm}, most of them derive and qualitatively confirm the one-loop correction by other methods, although quantitatively has not been settled yet in the community. Despite of that, our claim \cite{Kristiano:2022maq} was criticized by the author of \cite{Riotto:2023hoz} who attempted to evaluate the same correction with a different method. In case of sharp transition of the second SR parameter, we have shown in \cite{Kristiano:2023scm} that after more careful and proper calculation, one can reproduce our result even if the method proposed in \cite{Riotto:2023hoz} is used. In case the second SR parameter evolves continuously with function \eqref{etapsmooth}, the bispectrum is suppressed by a SR parameter $\eta_V$ \cite{Riotto:2023gpm}. In \cite{Kristiano:2024vst}, we have argued that suppression on the bispectrum and one-loop correction occur only for a specific smooth transition that satisfies Wands duality condition \cite{Wands:1998yp}.

Indeed, there are issues related to regularization and renormalization of the divergence. In \cite{Kristiano:2022maq}, we show that constraint \eqref{bound} corresponds to the requirement to perform the renormalization order by order in perturbation theory. Another method for performing regularization, renormalization, and resummation of loop corrections is proposed in \cite{Choudhury:2023vuj, Choudhury:2023jlt, Choudhury:2023rks}, where the sound speed of the curvature perturbation is introduced within the effective field theory of inflation framework. Numerical integration of the one-loop correction has been performed in \cite{Franciolini:2023lgy, Davies:2023hhn}, confirming our analytical result in the limit of the sharp transition. Also, Ref. \cite{Franciolini:2023lgy} pointed out that a dip in the power spectrum \eqref{kdip} is a tree-level artifact, which disappears after adding a one-loop correction. One-loop correction to the power spectrum around the peak scale has been investigated in \cite{Fumagalli:2023loc}. Specifically, for an inflationary potential with oscillatory features, it has been calculated in \cite{Inomata:2022yte, Caravano:2024tlp}.

Moreover, the $\delta N$ formalism and stochastic inflation in the presence of a sharp transition of the second SR parameter have been investigated in \cite{Mishra:2023lhe, Jackson:2023obv}. The one-loop correction has also been derived from the $\delta N$ formalism in \cite{Firouzjahi:2023ahg, Iacconi:2023ggt}. The extension from USR to constrant-roll period has been derived in \cite{Motohashi:2023syh}, which yields a constraint involving a constant second SR parameter and small-scale power spectrum. In the limit $\abs{\epsilon_2} \rightarrow \infty$, bispectrum and one-loop correction has been investigated in \cite{Tasinato:2023ioq, Tasinato:2023ukp}, based on a method that approximates the curvature perturbation in the presence of a very short period where the SR approximation is violated \cite{Tasinato:2020vdk}. In addition to cubic self-interaction, quartic self-interaction \cite{Jarnhus:2007ia, Arroja:2008ga} in the Hamiltonian can also contribute to the one-loop correction, in first-order perturbation theory. Its contribution has been investigated, either by direct expansion of the quartic derivative of the potential \cite{Maity:2023qzw} or by an effective field theory approach \cite{Akhshik:2015nfa, Firouzjahi:2023aum}. 

However, there are two works that completely disagree with our conclusions \cite{Fumagalli:2023hpa, Tada:2023rgp}. In their approach, instead of performing field redefinition, they calculate a one-loop correction directly from third-order action with total time derivative terms and terms proportional to the equation of motion, as shown in \eqref{S3}. Currently, they have been criticized by \cite{Firouzjahi:2023bkt}. Other perspectives on the effect of the total time derivative on the correlation function have been presented in \cite{Braglia:2024zsl, Kawaguchi:2024lsw}. We hope to solve this discrepancy by pointing out what is missing in \cite{Fumagalli:2023hpa, Tada:2023rgp}. Beyond the power spectrum, one-loop correction to the bispectrum has been explored in \cite{Firouzjahi:2024psd}.

While we have been performing computation in the comoving gauge, some authors do so adopting the flat-slicing gauge. In this gauge, perturbation is represented by the inflaton's fluctuation $\delta\phi$, with vanishing curvature perturbation. Prior to our work, the quantum correction to $\delta\phi$ was investigated in \cite{Syu:2019uwx, Cheng:2021lif} and followed up in \cite{Cheng:2023ikq}, which is consistent with our claim. The effect of a smooth transition of the second SR parameter in the flat-slicing gauge has been explored in \cite{Saburov:2024und, Ballesteros:2024zdp}. However, there is a result \cite{Inomata:2024lud} that is inconsistent with our claim due to inappropriate order counts.
\subsection{Summary}
In this chapter, we have explained the mechanism to generate large primordial fluctuation in single-field inflation. In order to amplify the curvature perturbation, violation of the SR approximation is needed. We have discussed the simplest example of an inflation model that violates the SR approximation, called USR inflation. We have shown that USR inflation leads to growth of the curvature perturbation outside the horizon. The consequence is that USR inflation cannot explain the required e-folds number to solve the horizon problem.

Nevertheless, one can utilize the properties of the USR period for another purpose. If the USR period is sandwiched by two SR periods, the curvature perturbation grows temporary during the USR period. At the end of inflation, the power spectrum of the curvature perturbation is amplified on small scales corresponding to the time scale of the USR period, while keeping the small amplitude on large scale consistent with CMB observations.

Moreover, we have explained the nonlinear evolution of the curvature perturbation induced by cubic self-interaction, which yields three-point functions or bispectrum. The cubic self-interaction consists of bulk interaction and field redefinition. In USR inflation, only the field redefinition contributes to the bispectrum. When a transition from USR to SR period exists, the bulk interaction that is proportional to the time derivative of the second SR parameter becomes important. If the transition is sharp, the nonlinear parameter of the bispectrum has a $\mathcal{O}(1)$ value. This means that the bispectrum is large enough, which indicates that the nonlinear effect cannot be neglected. At the same time, such a nonlinear effect gives backreaction to the power spectrum as a one-loop correction that leads to a constraint on the small-scale power spectrum associated with transition of the second SR parameter.

\acknowledgments
J.~K. acknowledges the support from JSPS KAKENHI Grants No.~22KJ1006 and No.~22J20289. J.~K. is also supported by the Global Science Graduate Course (GSGC) program of The University of Tokyo. J.~Y. is supported by JSPS KAKENHI Grant No.~20H05639. 

\bibliographystyle{jhep}
\bibliography{reference}

\providecommand{\noopsort}[1]{}\providecommand{\singleletter}[1]{#1}%

\providecommand{\href}[2]{#2}\begingroup\raggedright\begin{thebibliography}{100}

\bibitem{Zel:1967}
Y.B.~{Zel'dovich} and I.D.~{Novikov}, \emph{{The Hypothesis of Cores Retarded during Expansion and the Hot Cosmological Model}}, {\emph{Sov. Astron.} {\bfseries 10} (1967) 602}.

\bibitem{Hawking:1971ei}
S.~Hawking, \emph{{Gravitationally collapsed objects of very low mass}}, {\emph{Mon. Not. Roy. Astron. Soc.} {\bfseries 152} (1971) 75}.

\bibitem{Carr:1974nx}
B.J.~Carr and S.W.~Hawking, \emph{{Black holes in the early Universe}}, {\emph{Mon. Not. Roy. Astron. Soc.} {\bfseries 168} (1974) 399}.

\bibitem{Hawking:1974rv}
S.W.~Hawking, \emph{{Black hole explosions}}, \href{https://doi.org/10.1038/248030a0}{\emph{Nature} {\bfseries 248} (1974) 30}.

\bibitem{Starobinsky:1980te}
A.A.~Starobinsky, \emph{{A New Type of Isotropic Cosmological Models Without Singularity}}, \href{https://doi.org/10.1016/0370-2693(80)90670-X}{\emph{Phys. Lett. B} {\bfseries 91} (1980) 99}.

\bibitem{Sato:1980yn}
K.~Sato, \emph{{First Order Phase Transition of a Vacuum and Expansion of the Universe}}, {\emph{Mon. Not. Roy. Astron. Soc.} {\bfseries 195} (1981) 467}.

\bibitem{Guth:1980zm}
A.H.~Guth, \emph{{The Inflationary Universe: A Possible Solution to the Horizon and Flatness Problems}}, \href{https://doi.org/10.1103/PhysRevD.23.347}{\emph{Phys. Rev. D} {\bfseries 23} (1981) 347}.

\bibitem{Sato:2015dga}
K.~Sato and J.~Yokoyama, \emph{{Inflationary cosmology: First 30+ years}}, \href{https://doi.org/10.1142/S0218271815300256}{\emph{Int. J. Mod. Phys. D} {\bfseries 24} (2015) 1530025}.

\bibitem{Kristiano:2022maq}
J.~Kristiano and J.~Yokoyama, \emph{{Constraining Primordial Black Hole Formation from Single-Field Inflation}}, \href{https://doi.org/10.1103/PhysRevLett.132.221003}{\emph{Phys. Rev. Lett.} {\bfseries 132} (2024) 221003} [\href{https://arxiv.org/abs/2211.03395}{{\ttfamily 2211.03395}}].

\bibitem{Kristiano:2023scm}
J.~Kristiano and J.~Yokoyama, \emph{{Note on the bispectrum and one-loop corrections in single-field inflation with primordial black hole formation}}, \href{https://doi.org/10.1103/PhysRevD.109.103541}{\emph{Phys. Rev. D} {\bfseries 109} (2024) 103541} [\href{https://arxiv.org/abs/2303.00341}{{\ttfamily 2303.00341}}].

\bibitem{Chluba:2012we}
J.~Chluba, A.L.~Erickcek and I.~Ben-Dayan, \emph{{Probing the inflaton: Small-scale power spectrum constraints from measurements of the CMB energy spectrum}}, \href{https://doi.org/10.1088/0004-637X/758/2/76}{\emph{Astrophys. J.} {\bfseries 758} (2012) 76} [\href{https://arxiv.org/abs/1203.2681}{{\ttfamily 1203.2681}}].

\bibitem{Chluba:2019nxa}
J.~Chluba et~al., \emph{{New horizons in cosmology with spectral distortions of the cosmic microwave background}}, \href{https://doi.org/10.1007/s10686-021-09729-5}{\emph{Exper. Astron.} {\bfseries 51} (2021) 1515} [\href{https://arxiv.org/abs/1909.01593}{{\ttfamily 1909.01593}}].

\bibitem{Nakama:2014vla}
T.~Nakama, T.~Suyama and J.~Yokoyama, \emph{{Reheating the Universe Once More: The Dissipation of Acoustic Waves as a Novel Probe of Primordial Inhomogeneities on Even Smaller Scales}}, \href{https://doi.org/10.1103/PhysRevLett.113.061302}{\emph{Phys. Rev. Lett.} {\bfseries 113} (2014) 061302} [\href{https://arxiv.org/abs/1403.5407}{{\ttfamily 1403.5407}}].

\bibitem{Jeong:2014gna}
D.~Jeong, J.~Pradler, J.~Chluba and M.~Kamionkowski, \emph{{Silk damping at a redshift of a billion: a new limit on small-scale adiabatic perturbations}}, \href{https://doi.org/10.1103/PhysRevLett.113.061301}{\emph{Phys. Rev. Lett.} {\bfseries 113} (2014) 061301} [\href{https://arxiv.org/abs/1403.3697}{{\ttfamily 1403.3697}}].

\bibitem{Inomata:2016uip}
K.~Inomata, M.~Kawasaki and Y.~Tada, \emph{{Revisiting constraints on small scale perturbations from big-bang nucleosynthesis}}, \href{https://doi.org/10.1103/PhysRevD.94.043527}{\emph{Phys. Rev. D} {\bfseries 94} (2016) 043527} [\href{https://arxiv.org/abs/1605.04646}{{\ttfamily 1605.04646}}].

\bibitem{Nakama:2017qac}
T.~Nakama, T.~Suyama, K.~Kohri and N.~Hiroshima, \emph{{Constraints on small-scale primordial power by annihilation signals from extragalactic dark matter minihalos}}, \href{https://doi.org/10.1103/PhysRevD.97.023539}{\emph{Phys. Rev. D} {\bfseries 97} (2018) 023539} [\href{https://arxiv.org/abs/1712.08820}{{\ttfamily 1712.08820}}].

\bibitem{Kawasaki:2021yek}
M.~Kawasaki, H.~Nakatsuka and K.~Nakayama, \emph{{Constraints on small-scale primordial density fluctuation from cosmic microwave background through dark matter annihilation}}, \href{https://doi.org/10.1088/1475-7516/2022/03/061}{\emph{JCAP} {\bfseries 03} (2022) 061} [\href{https://arxiv.org/abs/2110.12620}{{\ttfamily 2110.12620}}].

\bibitem{Kimura:2021sqz}
R.~Kimura, T.~Suyama, M.~Yamaguchi and Y.-L.~Zhang, \emph{{Reconstruction of Primordial Power Spectrum of curvature perturbation from the merger rate of Primordial Black Hole Binaries}}, \href{https://doi.org/10.1088/1475-7516/2021/04/031}{\emph{JCAP} {\bfseries 04} (2021) 031} [\href{https://arxiv.org/abs/2102.05280}{{\ttfamily 2102.05280}}].

\bibitem{Wang:2022nml}
X.~Wang, Y.-l.~Zhang, R.~Kimura and M.~Yamaguchi, \emph{{Reconstruction of power spectrum of primordial curvature perturbations on small scales from primordial black hole binaries scenario of LIGO/VIRGO detection}}, \href{https://doi.org/10.1007/s11433-023-2091-x}{\emph{Sci. China Phys. Mech. Astron.} {\bfseries 66} (2023) 260462} [\href{https://arxiv.org/abs/2209.12911}{{\ttfamily 2209.12911}}].

\bibitem{Planck:2018nkj}
{\scshape Planck} collaboration, \emph{{Planck 2018 results. I. Overview and the cosmological legacy of Planck}}, \href{https://doi.org/10.1051/0004-6361/201833880}{\emph{Astron. Astrophys.} {\bfseries 641} (2020) A1} [\href{https://arxiv.org/abs/1807.06205}{{\ttfamily 1807.06205}}].

\bibitem{Planck:2018jri}
{\scshape Planck} collaboration, \emph{{Planck 2018 results. X. Constraints on inflation}}, \href{https://doi.org/10.1051/0004-6361/201833887}{\emph{Astron. Astrophys.} {\bfseries 641} (2020) A10} [\href{https://arxiv.org/abs/1807.06211}{{\ttfamily 1807.06211}}].

\bibitem{Planck:2019kim}
{\scshape Planck} collaboration, \emph{{Planck 2018 results. IX. Constraints on primordial non-Gaussianity}}, \href{https://doi.org/10.1051/0004-6361/201935891}{\emph{Astron. Astrophys.} {\bfseries 641} (2020) A9} [\href{https://arxiv.org/abs/1905.05697}{{\ttfamily 1905.05697}}].

\bibitem{Yokoyama:1998rw}
J.~Yokoyama, \emph{{Chaotic new inflation and primordial spectrum of adiabatic fluctuations}}, \href{https://doi.org/10.1103/PhysRevD.59.107303}{\emph{Phys. Rev. D} {\bfseries 59} (1999) 107303}.

\bibitem{Saito:2008em}
R.~Saito, J.~Yokoyama and R.~Nagata, \emph{{Single-field inflation, anomalous enhancement of superhorizon fluctuations, and non-Gaussianity in primordial black hole formation}}, \href{https://doi.org/10.1088/1475-7516/2008/06/024}{\emph{JCAP} {\bfseries 06} (2008) 024} [\href{https://arxiv.org/abs/0804.3470}{{\ttfamily 0804.3470}}].

\bibitem{Kinney:1997ne}
W.H.~Kinney, \emph{{A Hamilton-Jacobi approach to nonslow roll inflation}}, \href{https://doi.org/10.1103/PhysRevD.56.2002}{\emph{Phys. Rev. D} {\bfseries 56} (1997) 2002} [\href{https://arxiv.org/abs/hep-ph/9702427}{{\ttfamily hep-ph/9702427}}].

\bibitem{Inoue:2001zt}
J.~Yokoyama and S.~Inoue, \emph{{Curvature perturbation at the local extremum of the inflaton's potential}}, \href{https://doi.org/10.1016/S0370-2693(01)01369-7}{\emph{Phys. Lett. B} {\bfseries 524} (2002) 15} [\href{https://arxiv.org/abs/hep-ph/0104083}{{\ttfamily hep-ph/0104083}}].

\bibitem{Kinney:2005vj}
W.H.~Kinney, \emph{{Horizon crossing and inflation with large eta}}, \href{https://doi.org/10.1103/PhysRevD.72.023515}{\emph{Phys. Rev. D} {\bfseries 72} (2005) 023515} [\href{https://arxiv.org/abs/gr-qc/0503017}{{\ttfamily gr-qc/0503017}}].

\bibitem{Martin:2012pe}
J.~Martin, H.~Motohashi and T.~Suyama, \emph{{Ultra Slow-Roll Inflation and the non-Gaussianity Consistency Relation}}, \href{https://doi.org/10.1103/PhysRevD.87.023514}{\emph{Phys. Rev. D} {\bfseries 87} (2013) 023514} [\href{https://arxiv.org/abs/1211.0083}{{\ttfamily 1211.0083}}].

\bibitem{Motohashi:2017kbs}
H.~Motohashi and W.~Hu, \emph{{Primordial Black Holes and Slow-Roll Violation}}, \href{https://doi.org/10.1103/PhysRevD.96.063503}{\emph{Phys. Rev. D} {\bfseries 96} (2017) 063503} [\href{https://arxiv.org/abs/1706.06784}{{\ttfamily 1706.06784}}].

\bibitem{Motohashi:2014ppa}
H.~Motohashi, A.A.~Starobinsky and J.~Yokoyama, \emph{{Inflation with a constant rate of roll}}, \href{https://doi.org/10.1088/1475-7516/2015/09/018}{\emph{JCAP} {\bfseries 09} (2015) 018} [\href{https://arxiv.org/abs/1411.5021}{{\ttfamily 1411.5021}}].

\bibitem{Motohashi:2017aob}
H.~Motohashi and A.A.~Starobinsky, \emph{{Constant-roll inflation: confrontation with recent observational data}}, \href{https://doi.org/10.1209/0295-5075/117/39001}{\emph{EPL} {\bfseries 117} (2017) 39001} [\href{https://arxiv.org/abs/1702.05847}{{\ttfamily 1702.05847}}].

\bibitem{Motohashi:2019rhu}
H.~Motohashi, S.~Mukohyama and M.~Oliosi, \emph{{Constant Roll and Primordial Black Holes}}, \href{https://doi.org/10.1088/1475-7516/2020/03/002}{\emph{JCAP} {\bfseries 03} (2020) 002} [\href{https://arxiv.org/abs/1910.13235}{{\ttfamily 1910.13235}}].

\bibitem{Ivanov:1994pa}
P.~Ivanov, P.~Naselsky and I.~Novikov, \emph{{Inflation and primordial black holes as dark matter}}, \href{https://doi.org/10.1103/PhysRevD.50.7173}{\emph{Phys. Rev. D} {\bfseries 50} (1994) 7173}.

\bibitem{Polarski:1995jg}
D.~Polarski and A.A.~Starobinsky, \emph{{Semiclassicality and decoherence of cosmological perturbations}}, \href{https://doi.org/10.1088/0264-9381/13/3/006}{\emph{Class. Quant. Grav.} {\bfseries 13} (1996) 377} [\href{https://arxiv.org/abs/gr-qc/9504030}{{\ttfamily gr-qc/9504030}}].

\bibitem{Cai:2018dkf}
Y.-F.~Cai, X.~Chen, M.H.~Namjoo, M.~Sasaki, D.-G.~Wang and Z.~Wang, \emph{{Revisiting non-Gaussianity from non-attractor inflation models}}, \href{https://doi.org/10.1088/1475-7516/2018/05/012}{\emph{JCAP} {\bfseries 05} (2018) 012} [\href{https://arxiv.org/abs/1712.09998}{{\ttfamily 1712.09998}}].

\bibitem{Starobinsky:1992ts}
A.A.~Starobinsky, \emph{{Spectrum of adiabatic perturbations in the universe when there are singularities in the inflation potential}}, {\emph{JETP Lett.} {\bfseries 55} (1992) 489}.

\bibitem{Leach:2001zf}
S.M.~Leach, M.~Sasaki, D.~Wands and A.R.~Liddle, \emph{{Enhancement of superhorizon scale inflationary curvature perturbations}}, \href{https://doi.org/10.1103/PhysRevD.64.023512}{\emph{Phys. Rev. D} {\bfseries 64} (2001) 023512} [\href{https://arxiv.org/abs/astro-ph/0101406}{{\ttfamily astro-ph/0101406}}].

\bibitem{Byrnes:2018txb}
C.T.~Byrnes, P.S.~Cole and S.P.~Patil, \emph{{Steepest growth of the power spectrum and primordial black holes}}, \href{https://doi.org/10.1088/1475-7516/2019/06/028}{\emph{JCAP} {\bfseries 06} (2019) 028} [\href{https://arxiv.org/abs/1811.11158}{{\ttfamily 1811.11158}}].

\bibitem{Liu:2020oqe}
J.~Liu, Z.-K.~Guo and R.-G.~Cai, \emph{{Analytical approximation of the scalar spectrum in the ultraslow-roll inflationary models}}, \href{https://doi.org/10.1103/PhysRevD.101.083535}{\emph{Phys. Rev. D} {\bfseries 101} (2020) 083535} [\href{https://arxiv.org/abs/2003.02075}{{\ttfamily 2003.02075}}].

\bibitem{Tasinato:2020vdk}
G.~Tasinato, \emph{{An analytic approach to non-slow-roll inflation}}, \href{https://doi.org/10.1103/PhysRevD.103.023535}{\emph{Phys. Rev. D} {\bfseries 103} (2021) 023535} [\href{https://arxiv.org/abs/2012.02518}{{\ttfamily 2012.02518}}].

\bibitem{Karam:2022nym}
A.~Karam, N.~Koivunen, E.~Tomberg, V.~Vaskonen and H.~Veerm\"ae, \emph{{Anatomy of single-field inflationary models for primordial black holes}}, \href{https://doi.org/10.1088/1475-7516/2023/03/013}{\emph{JCAP} {\bfseries 03} (2023) 013} [\href{https://arxiv.org/abs/2205.13540}{{\ttfamily 2205.13540}}].

\bibitem{Pi:2022zxs}
S.~Pi and J.~Wang, \emph{{Primordial black hole formation in Starobinsky's linear potential model}}, \href{https://doi.org/10.1088/1475-7516/2023/06/018}{\emph{JCAP} {\bfseries 06} (2023) 018} [\href{https://arxiv.org/abs/2209.14183}{{\ttfamily 2209.14183}}].

\bibitem{Domenech:2023dxx}
G.~Dom\`enech, G.~Vargas and T.~Vargas, \emph{{An exact model for enhancing/suppressing primordial fluctuations}}, \href{https://doi.org/10.1088/1475-7516/2024/03/002}{\emph{JCAP} {\bfseries 03} (2024) 002} [\href{https://arxiv.org/abs/2309.05750}{{\ttfamily 2309.05750}}].

\bibitem{Maldacena:2002vr}
J.M.~Maldacena, \emph{{Non-Gaussian features of primordial fluctuations in single field inflationary models}}, \href{https://doi.org/10.1088/1126-6708/2003/05/013}{\emph{JHEP} {\bfseries 05} (2003) 013} [\href{https://arxiv.org/abs/astro-ph/0210603}{{\ttfamily astro-ph/0210603}}].

\bibitem{Arroja:2011yj}
F.~Arroja and T.~Tanaka, \emph{{A note on the role of the boundary terms for the non-Gaussianity in general k-inflation}}, \href{https://doi.org/10.1088/1475-7516/2011/05/005}{\emph{JCAP} {\bfseries 05} (2011) 005} [\href{https://arxiv.org/abs/1103.1102}{{\ttfamily 1103.1102}}].

\bibitem{Burrage:2011hd}
C.~Burrage, R.H.~Ribeiro and D.~Seery, \emph{{Large slow-roll corrections to the bispectrum of noncanonical inflation}}, \href{https://doi.org/10.1088/1475-7516/2011/07/032}{\emph{JCAP} {\bfseries 07} (2011) 032} [\href{https://arxiv.org/abs/1103.4126}{{\ttfamily 1103.4126}}].

\bibitem{Creminelli:2004yq}
P.~Creminelli and M.~Zaldarriaga, \emph{{Single field consistency relation for the 3-point function}}, \href{https://doi.org/10.1088/1475-7516/2004/10/006}{\emph{JCAP} {\bfseries 10} (2004) 006} [\href{https://arxiv.org/abs/astro-ph/0407059}{{\ttfamily astro-ph/0407059}}].

\bibitem{Namjoo:2012aa}
M.H.~Namjoo, H.~Firouzjahi and M.~Sasaki, \emph{{Violation of non-Gaussianity consistency relation in a single field inflationary model}}, \href{https://doi.org/10.1209/0295-5075/101/39001}{\emph{EPL} {\bfseries 101} (2013) 39001} [\href{https://arxiv.org/abs/1210.3692}{{\ttfamily 1210.3692}}].

\bibitem{Bravo:2017wyw}
R.~Bravo, S.~Mooij, G.A.~Palma and B.~Pradenas, \emph{{A generalized non-Gaussian consistency relation for single field inflation}}, \href{https://doi.org/10.1088/1475-7516/2018/05/024}{\emph{JCAP} {\bfseries 05} (2018) 024} [\href{https://arxiv.org/abs/1711.02680}{{\ttfamily 1711.02680}}].

\bibitem{Finelli:2017fml}
B.~Finelli, G.~Goon, E.~Pajer and L.~Santoni, \emph{{Soft Theorems For Shift-Symmetric Cosmologies}}, \href{https://doi.org/10.1103/PhysRevD.97.063531}{\emph{Phys. Rev. D} {\bfseries 97} (2018) 063531} [\href{https://arxiv.org/abs/1711.03737}{{\ttfamily 1711.03737}}].

\bibitem{Cai:2016ngx}
Y.-F.~Cai, J.-O.~Gong, D.-G.~Wang and Z.~Wang, \emph{{Features from the non-attractor beginning of inflation}}, \href{https://doi.org/10.1088/1475-7516/2016/10/017}{\emph{JCAP} {\bfseries 10} (2016) 017} [\href{https://arxiv.org/abs/1607.07872}{{\ttfamily 1607.07872}}].

\bibitem{Chen:2013eea}
X.~Chen, H.~Firouzjahi, E.~Komatsu, M.H.~Namjoo and M.~Sasaki, \emph{{In-in and $\delta N$ calculations of the bispectrum from non-attractor single-field inflation}}, \href{https://doi.org/10.1088/1475-7516/2013/12/039}{\emph{JCAP} {\bfseries 12} (2013) 039} [\href{https://arxiv.org/abs/1308.5341}{{\ttfamily 1308.5341}}].

\bibitem{Davies:2021loj}
M.W.~Davies, P.~Carrilho and D.J.~Mulryne, \emph{{Non-Gaussianity in inflationary scenarios for primordial black holes}}, \href{https://doi.org/10.1088/1475-7516/2022/06/019}{\emph{JCAP} {\bfseries 06} (2022) 019} [\href{https://arxiv.org/abs/2110.08189}{{\ttfamily 2110.08189}}].

\bibitem{Kristiano:2024vst}
J.~Kristiano and J.~Yokoyama, \emph{{Comparing sharp and smooth transitions of the second slow-roll parameter in single-field inflation}},  \href{https://arxiv.org/abs/2405.12145}{{\ttfamily 2405.12145}}.

\bibitem{Wands:1998yp}
D.~Wands, \emph{{Duality invariance of cosmological perturbation spectra}}, \href{https://doi.org/10.1103/PhysRevD.60.023507}{\emph{Phys. Rev. D} {\bfseries 60} (1999) 023507} [\href{https://arxiv.org/abs/gr-qc/9809062}{{\ttfamily gr-qc/9809062}}].

\bibitem{Namjoo:2023rhq}
M.H.~Namjoo, \emph{{One consistency relation for all single-field inflationary models}}, \href{https://doi.org/10.1088/1475-7516/2024/05/041}{\emph{JCAP} {\bfseries 05} (2024) 041} [\href{https://arxiv.org/abs/2311.12777}{{\ttfamily 2311.12777}}].

\bibitem{Balaji:2022zur}
S.~Balaji, H.V.~Ragavendra, S.K.~Sethi, J.~Silk and L.~Sriramkumar, \emph{{Observing Nulling of Primordial Correlations via the 21-cm Signal}}, \href{https://doi.org/10.1103/PhysRevLett.129.261301}{\emph{Phys. Rev. Lett.} {\bfseries 129} (2022) 261301} [\href{https://arxiv.org/abs/2206.06386}{{\ttfamily 2206.06386}}].

\bibitem{Namjoo:2024ufv}
M.H.~Namjoo and B.~Nikbakht, \emph{{Non-Gaussianity consistency relations and their consequences for the peaks}},  \href{https://arxiv.org/abs/2401.12958}{{\ttfamily 2401.12958}}.

\bibitem{Firouzjahi:2023xke}
H.~Firouzjahi and A.~Riotto, \emph{{Sign of non-Gaussianity and the primordial black holes abundance}}, \href{https://doi.org/10.1103/PhysRevD.108.123504}{\emph{Phys. Rev. D} {\bfseries 108} (2023) 123504} [\href{https://arxiv.org/abs/2309.10536}{{\ttfamily 2309.10536}}].

\bibitem{Kawasaki:1997ju}
M.~Kawasaki, N.~Sugiyama and T.~Yanagida, \emph{{Primordial black hole formation in a double inflation model in supergravity}}, \href{https://doi.org/10.1103/PhysRevD.57.6050}{\emph{Phys. Rev. D} {\bfseries 57} (1998) 6050} [\href{https://arxiv.org/abs/hep-ph/9710259}{{\ttfamily hep-ph/9710259}}].

\bibitem{Kawasaki:1998vx}
M.~Kawasaki and T.~Yanagida, \emph{{Primordial black hole formation in supergravity}}, \href{https://doi.org/10.1103/PhysRevD.59.043512}{\emph{Phys. Rev. D} {\bfseries 59} (1999) 043512} [\href{https://arxiv.org/abs/hep-ph/9807544}{{\ttfamily hep-ph/9807544}}].

\bibitem{Kawasaki:2016pql}
M.~Kawasaki, A.~Kusenko, Y.~Tada and T.T.~Yanagida, \emph{{Primordial black holes as dark matter in supergravity inflation models}}, \href{https://doi.org/10.1103/PhysRevD.94.083523}{\emph{Phys. Rev. D} {\bfseries 94} (2016) 083523} [\href{https://arxiv.org/abs/1606.07631}{{\ttfamily 1606.07631}}].

\bibitem{Gao:2018pvq}
T.-J.~Gao and Z.-K.~Guo, \emph{{Primordial Black Hole Production in Inflationary Models of Supergravity with a Single Chiral Superfield}}, \href{https://doi.org/10.1103/PhysRevD.98.063526}{\emph{Phys. Rev. D} {\bfseries 98} (2018) 063526} [\href{https://arxiv.org/abs/1806.09320}{{\ttfamily 1806.09320}}].

\bibitem{Nanopoulos:2020nnh}
D.V.~Nanopoulos, V.C.~Spanos and I.D.~Stamou, \emph{{Primordial Black Holes from No-Scale Supergravity}}, \href{https://doi.org/10.1103/PhysRevD.102.083536}{\emph{Phys. Rev. D} {\bfseries 102} (2020) 083536} [\href{https://arxiv.org/abs/2008.01457}{{\ttfamily 2008.01457}}].

\bibitem{Wu:2021zta}
L.~Wu, Y.~Gong and T.~Li, \emph{{Primordial black holes and secondary gravitational waves from string inspired general no-scale supergravity}}, \href{https://doi.org/10.1103/PhysRevD.104.123544}{\emph{Phys. Rev. D} {\bfseries 104} (2021) 123544} [\href{https://arxiv.org/abs/2105.07694}{{\ttfamily 2105.07694}}].

\bibitem{Stamou:2021qdk}
I.D.~Stamou, \emph{{Mechanisms of producing primordial black holes by breaking the $SU(2, 1)/SU(2)\times U(1)$ symmetry}}, \href{https://doi.org/10.1103/PhysRevD.103.083512}{\emph{Phys. Rev. D} {\bfseries 103} (2021) 083512} [\href{https://arxiv.org/abs/2104.08654}{{\ttfamily 2104.08654}}].

\bibitem{Cheng:2018yyr}
S.-L.~Cheng, W.~Lee and K.-W.~Ng, \emph{{Primordial black holes and associated gravitational waves in axion monodromy inflation}}, \href{https://doi.org/10.1088/1475-7516/2018/07/001}{\emph{JCAP} {\bfseries 07} (2018) 001} [\href{https://arxiv.org/abs/1801.09050}{{\ttfamily 1801.09050}}].

\bibitem{Ballesteros:2019hus}
G.~Ballesteros, J.~Rey and F.~Rompineve, \emph{{Detuning primordial black hole dark matter with early matter domination and axion monodromy}}, \href{https://doi.org/10.1088/1475-7516/2020/06/014}{\emph{JCAP} {\bfseries 06} (2020) 014} [\href{https://arxiv.org/abs/1912.01638}{{\ttfamily 1912.01638}}].

\bibitem{Pi:2017gih}
S.~Pi, Y.-l.~Zhang, Q.-G.~Huang and M.~Sasaki, \emph{{Scalaron from $R^2$-gravity as a heavy field}}, \href{https://doi.org/10.1088/1475-7516/2018/05/042}{\emph{JCAP} {\bfseries 05} (2018) 042} [\href{https://arxiv.org/abs/1712.09896}{{\ttfamily 1712.09896}}].

\bibitem{Dalianis:2019asr}
I.~Dalianis and G.~Tringas, \emph{{Primordial black hole remnants as dark matter produced in thermal, matter, and runaway-quintessence postinflationary scenarios}}, \href{https://doi.org/10.1103/PhysRevD.100.083512}{\emph{Phys. Rev. D} {\bfseries 100} (2019) 083512} [\href{https://arxiv.org/abs/1905.01741}{{\ttfamily 1905.01741}}].

\bibitem{Dalianis:2018frf}
I.~Dalianis, A.~Kehagias and G.~Tringas, \emph{{Primordial black holes from \ensuremath{\alpha}-attractors}}, \href{https://doi.org/10.1088/1475-7516/2019/01/037}{\emph{JCAP} {\bfseries 01} (2019) 037} [\href{https://arxiv.org/abs/1805.09483}{{\ttfamily 1805.09483}}].

\bibitem{Mahbub:2019uhl}
R.~Mahbub, \emph{{Primordial black hole formation in inflationary $\alpha$-attractor models}}, \href{https://doi.org/10.1103/PhysRevD.101.023533}{\emph{Phys. Rev. D} {\bfseries 101} (2020) 023533} [\href{https://arxiv.org/abs/1910.10602}{{\ttfamily 1910.10602}}].

\bibitem{Fu:2022ypp}
C.~Fu and S.-J.~Wang, \emph{{Primordial black holes and induced gravitational waves from double-pole inflation}}, \href{https://doi.org/10.1088/1475-7516/2023/06/012}{\emph{JCAP} {\bfseries 06} (2023) 012} [\href{https://arxiv.org/abs/2211.03523}{{\ttfamily 2211.03523}}].

\bibitem{Cicoli:2018asa}
M.~Cicoli, V.A.~Diaz and F.G.~Pedro, \emph{{Primordial Black Holes from String Inflation}}, \href{https://doi.org/10.1088/1475-7516/2018/06/034}{\emph{JCAP} {\bfseries 06} (2018) 034} [\href{https://arxiv.org/abs/1803.02837}{{\ttfamily 1803.02837}}].

\bibitem{Ozsoy:2018flq}
O.~\"Ozsoy, S.~Parameswaran, G.~Tasinato and I.~Zavala, \emph{{Mechanisms for Primordial Black Hole Production in String Theory}}, \href{https://doi.org/10.1088/1475-7516/2018/07/005}{\emph{JCAP} {\bfseries 07} (2018) 005} [\href{https://arxiv.org/abs/1803.07626}{{\ttfamily 1803.07626}}].

\bibitem{Cicoli:2022sih}
M.~Cicoli, F.G.~Pedro and N.~Pedron, \emph{{Secondary GWs and PBHs in string inflation: formation and detectability}}, \href{https://doi.org/10.1088/1475-7516/2022/08/030}{\emph{JCAP} {\bfseries 08} (2022) 030} [\href{https://arxiv.org/abs/2203.00021}{{\ttfamily 2203.00021}}].

\bibitem{Ezquiaga:2017fvi}
J.M.~Ezquiaga, J.~Garcia-Bellido and E.~Ruiz~Morales, \emph{{Primordial Black Hole production in Critical Higgs Inflation}}, \href{https://doi.org/10.1016/j.physletb.2017.11.039}{\emph{Phys. Lett. B} {\bfseries 776} (2018) 345} [\href{https://arxiv.org/abs/1705.04861}{{\ttfamily 1705.04861}}].

\bibitem{Ballesteros:2017fsr}
G.~Ballesteros and M.~Taoso, \emph{{Primordial black hole dark matter from single field inflation}}, \href{https://doi.org/10.1103/PhysRevD.97.023501}{\emph{Phys. Rev. D} {\bfseries 97} (2018) 023501} [\href{https://arxiv.org/abs/1709.05565}{{\ttfamily 1709.05565}}].

\bibitem{Drees:2019xpp}
M.~Drees and Y.~Xu, \emph{{Overshooting, Critical Higgs Inflation and Second Order Gravitational Wave Signatures}}, \href{https://doi.org/10.1140/epjc/s10052-021-08976-2}{\emph{Eur. Phys. J. C} {\bfseries 81} (2021) 182} [\href{https://arxiv.org/abs/1905.13581}{{\ttfamily 1905.13581}}].

\bibitem{Cheong:2019vzl}
D.Y.~Cheong, S.M.~Lee and S.C.~Park, \emph{{Primordial black holes in Higgs-$R^2$ inflation as the whole of dark matter}}, \href{https://doi.org/10.1088/1475-7516/2021/01/032}{\emph{JCAP} {\bfseries 01} (2021) 032} [\href{https://arxiv.org/abs/1912.12032}{{\ttfamily 1912.12032}}].

\bibitem{Rasanen:2018fom}
S.~Rasanen and E.~Tomberg, \emph{{Planck scale black hole dark matter from Higgs inflation}}, \href{https://doi.org/10.1088/1475-7516/2019/01/038}{\emph{JCAP} {\bfseries 01} (2019) 038} [\href{https://arxiv.org/abs/1810.12608}{{\ttfamily 1810.12608}}].

\bibitem{Garcia-Bellido:2017mdw}
J.~Garcia-Bellido and E.~Ruiz~Morales, \emph{{Primordial black holes from single field models of inflation}}, \href{https://doi.org/10.1016/j.dark.2017.09.007}{\emph{Phys. Dark Univ.} {\bfseries 18} (2017) 47} [\href{https://arxiv.org/abs/1702.03901}{{\ttfamily 1702.03901}}].

\bibitem{Ragavendra:2020sop}
H.V.~Ragavendra, P.~Saha, L.~Sriramkumar and J.~Silk, \emph{{Primordial black holes and secondary gravitational waves from ultraslow roll and punctuated inflation}}, \href{https://doi.org/10.1103/PhysRevD.103.083510}{\emph{Phys. Rev. D} {\bfseries 103} (2021) 083510} [\href{https://arxiv.org/abs/2008.12202}{{\ttfamily 2008.12202}}].

\bibitem{Mishra:2019pzq}
S.S.~Mishra and V.~Sahni, \emph{{Primordial Black Holes from a tiny bump/dip in the Inflaton potential}}, \href{https://doi.org/10.1088/1475-7516/2020/04/007}{\emph{JCAP} {\bfseries 04} (2020) 007} [\href{https://arxiv.org/abs/1911.00057}{{\ttfamily 1911.00057}}].

\bibitem{Atal:2019cdz}
V.~Atal, J.~Garriga and A.~Marcos-Caballero, \emph{{Primordial black hole formation with non-Gaussian curvature perturbations}}, \href{https://doi.org/10.1088/1475-7516/2019/09/073}{\emph{JCAP} {\bfseries 09} (2019) 073} [\href{https://arxiv.org/abs/1905.13202}{{\ttfamily 1905.13202}}].

\bibitem{Zheng:2021vda}
R.~Zheng, J.~Shi and T.~Qiu, \emph{{On primordial black holes and secondary gravitational waves generated from inflation with solo/multi-bumpy potential}}, \href{https://doi.org/10.1088/1674-1137/ac42bd}{\emph{Chin. Phys. C} {\bfseries 46} (2022) 045103} [\href{https://arxiv.org/abs/2106.04303}{{\ttfamily 2106.04303}}].

\bibitem{Wang:2021kbh}
Q.~Wang, Y.-C.~Liu, B.-Y.~Su and N.~Li, \emph{{Primordial black holes from the perturbations in the inflaton potential in peak theory}}, \href{https://doi.org/10.1103/PhysRevD.104.083546}{\emph{Phys. Rev. D} {\bfseries 104} (2021) 083546} [\href{https://arxiv.org/abs/2111.10028}{{\ttfamily 2111.10028}}].

\bibitem{Rezazadeh:2021clf}
K.~Rezazadeh, Z.~Teimoori, S.~Karimi and K.~Karami, \emph{{Non-Gaussianity and secondary gravitational waves from primordial black holes production in $\alpha $-attractor inflation}}, \href{https://doi.org/10.1140/epjc/s10052-022-10735-w}{\emph{Eur. Phys. J. C} {\bfseries 82} (2022) 758} [\href{https://arxiv.org/abs/2110.01482}{{\ttfamily 2110.01482}}].

\bibitem{Iacconi:2021ltm}
L.~Iacconi, H.~Assadullahi, M.~Fasiello and D.~Wands, \emph{{Revisiting small-scale fluctuations in \ensuremath{\alpha}-attractor models of inflation}}, \href{https://doi.org/10.1088/1475-7516/2022/06/007}{\emph{JCAP} {\bfseries 06} (2022) 007} [\href{https://arxiv.org/abs/2112.05092}{{\ttfamily 2112.05092}}].

\bibitem{Cai:2021zsp}
Y.-F.~Cai, X.-H.~Ma, M.~Sasaki, D.-G.~Wang and Z.~Zhou, \emph{{One small step for an inflaton, one giant leap for inflation: A novel non-Gaussian tail and primordial black holes}}, \href{https://doi.org/10.1016/j.physletb.2022.137461}{\emph{Phys. Lett. B} {\bfseries 834} (2022) 137461} [\href{https://arxiv.org/abs/2112.13836}{{\ttfamily 2112.13836}}].

\bibitem{Kefala:2020xsx}
K.~Kefala, G.P.~Kodaxis, I.D.~Stamou and N.~Tetradis, \emph{{Features of the inflaton potential and the power spectrum of cosmological perturbations}}, \href{https://doi.org/10.1103/PhysRevD.104.023506}{\emph{Phys. Rev. D} {\bfseries 104} (2021) 023506} [\href{https://arxiv.org/abs/2010.12483}{{\ttfamily 2010.12483}}].

\bibitem{Ng:2021hll}
K.-W.~Ng and Y.-P.~Wu, \emph{{Constant-rate inflation: primordial black holes from conformal weight transitions}}, \href{https://doi.org/10.1007/JHEP11(2021)076}{\emph{JHEP} {\bfseries 11} (2021) 076} [\href{https://arxiv.org/abs/2102.05620}{{\ttfamily 2102.05620}}].

\bibitem{Inomata:2021uqj}
K.~Inomata, E.~McDonough and W.~Hu, \emph{{Primordial black holes arise when the inflaton falls}}, \href{https://doi.org/10.1103/PhysRevD.104.123553}{\emph{Phys. Rev. D} {\bfseries 104} (2021) 123553} [\href{https://arxiv.org/abs/2104.03972}{{\ttfamily 2104.03972}}].

\bibitem{Inomata:2021tpx}
K.~Inomata, E.~McDonough and W.~Hu, \emph{{Amplification of primordial perturbations from the rise or fall of the inflaton}}, \href{https://doi.org/10.1088/1475-7516/2022/02/031}{\emph{JCAP} {\bfseries 02} (2022) 031} [\href{https://arxiv.org/abs/2110.14641}{{\ttfamily 2110.14641}}].

\bibitem{Kawaguchi:2023mgk}
R.~Kawaguchi, T.~Fujita and M.~Sasaki, \emph{{Highly asymmetric probability distribution from a finite-width upward step during inflation}}, \href{https://doi.org/10.1088/1475-7516/2023/11/021}{\emph{JCAP} {\bfseries 11} (2023) 021} [\href{https://arxiv.org/abs/2305.18140}{{\ttfamily 2305.18140}}].

\bibitem{Hertzberg:2017dkh}
M.P.~Hertzberg and M.~Yamada, \emph{{Primordial Black Holes from Polynomial Potentials in Single Field Inflation}}, \href{https://doi.org/10.1103/PhysRevD.97.083509}{\emph{Phys. Rev. D} {\bfseries 97} (2018) 083509} [\href{https://arxiv.org/abs/1712.09750}{{\ttfamily 1712.09750}}].

\bibitem{Ballesteros:2020qam}
G.~Ballesteros, J.~Rey, M.~Taoso and A.~Urbano, \emph{{Primordial black holes as dark matter and gravitational waves from single-field polynomial inflation}}, \href{https://doi.org/10.1088/1475-7516/2020/07/025}{\emph{JCAP} {\bfseries 07} (2020) 025} [\href{https://arxiv.org/abs/2001.08220}{{\ttfamily 2001.08220}}].

\bibitem{Kannike:2017bxn}
K.~Kannike, L.~Marzola, M.~Raidal and H.~Veerm\"ae, \emph{{Single Field Double Inflation and Primordial Black Holes}}, \href{https://doi.org/10.1088/1475-7516/2017/09/020}{\emph{JCAP} {\bfseries 09} (2017) 020} [\href{https://arxiv.org/abs/1705.06225}{{\ttfamily 1705.06225}}].

\bibitem{Di:2017ndc}
H.~Di and Y.~Gong, \emph{{Primordial black holes and second order gravitational waves from ultra-slow-roll inflation}}, \href{https://doi.org/10.1088/1475-7516/2018/07/007}{\emph{JCAP} {\bfseries 07} (2018) 007} [\href{https://arxiv.org/abs/1707.09578}{{\ttfamily 1707.09578}}].

\bibitem{Yokoyama:1998pt}
J.~Yokoyama, \emph{{Chaotic new inflation and formation of primordial black holes}}, \href{https://doi.org/10.1103/PhysRevD.58.083510}{\emph{Phys. Rev. D} {\bfseries 58} (1998) 083510} [\href{https://arxiv.org/abs/astro-ph/9802357}{{\ttfamily astro-ph/9802357}}].

\bibitem{Bugaev:2008bi}
E.~Bugaev and P.~Klimai, \emph{{Large curvature perturbations near horizon crossing in single-field inflation models}}, \href{https://doi.org/10.1103/PhysRevD.78.063515}{\emph{Phys. Rev. D} {\bfseries 78} (2008) 063515} [\href{https://arxiv.org/abs/0806.4541}{{\ttfamily 0806.4541}}].

\bibitem{Karam:2023haj}
A.~Karam, N.~Koivunen, E.~Tomberg, A.~Racioppi and H.~Veerm\"ae, \emph{{Primordial black holes and inflation from double-well potentials}}, \href{https://doi.org/10.1088/1475-7516/2023/09/002}{\emph{JCAP} {\bfseries 09} (2023) 002} [\href{https://arxiv.org/abs/2305.09630}{{\ttfamily 2305.09630}}].

\bibitem{Armendariz-Picon:1999hyi}
C.~Armendariz-Picon, T.~Damour and V.F.~Mukhanov, \emph{{k - inflation}}, \href{https://doi.org/10.1016/S0370-2693(99)00603-6}{\emph{Phys. Lett. B} {\bfseries 458} (1999) 209} [\href{https://arxiv.org/abs/hep-th/9904075}{{\ttfamily hep-th/9904075}}].

\bibitem{Kobayashi:2011nu}
T.~Kobayashi, M.~Yamaguchi and J.~Yokoyama, \emph{{Generalized G-inflation: Inflation with the most general second-order field equations}}, \href{https://doi.org/10.1143/PTP.126.511}{\emph{Prog. Theor. Phys.} {\bfseries 126} (2011) 511} [\href{https://arxiv.org/abs/1105.5723}{{\ttfamily 1105.5723}}].

\bibitem{Solbi:2021wbo}
M.~Solbi and K.~Karami, \emph{{Primordial black holes and induced gravitational waves in $k$-inflation}}, \href{https://doi.org/10.1088/1475-7516/2021/08/056}{\emph{JCAP} {\bfseries 08} (2021) 056} [\href{https://arxiv.org/abs/2102.05651}{{\ttfamily 2102.05651}}].

\bibitem{Solbi:2021rse}
M.~Solbi and K.~Karami, \emph{{Primordial black holes formation in the inflationary model with field-dependent kinetic term for quartic and natural potentials}}, \href{https://doi.org/10.1140/epjc/s10052-021-09690-9}{\emph{Eur. Phys. J. C} {\bfseries 81} (2021) 884} [\href{https://arxiv.org/abs/2106.02863}{{\ttfamily 2106.02863}}].

\bibitem{Teimoori:2021pte}
Z.~Teimoori, K.~Rezazadeh, M.A.~Rasheed and K.~Karami, \emph{Mechanism of primordial black holes production and secondary gravitational waves in $\alpha$-attractor galileon inflationary scenario}, \href{https://doi.org/10.1088/1475-7516/2021/10/018}{\emph{JCAP} {\bfseries 2021} (2021) 018} [\href{https://arxiv.org/abs/2107.07620}{{\ttfamily 2107.07620}}].

\bibitem{Heydari:2023xts}
S.~Heydari and K.~Karami, \emph{{Primordial black holes in non-canonical scalar field inflation driven by quartic potential in the presence of bump}}, \href{https://doi.org/10.1088/1475-7516/2024/02/047}{\emph{JCAP} {\bfseries 02} (2024) 047} [\href{https://arxiv.org/abs/2309.01239}{{\ttfamily 2309.01239}}].

\bibitem{Heydari:2023rmq}
S.~Heydari and K.~Karami, \emph{{Primordial black holes and secondary gravitational waves from generalized power-law non-canonical inflation with quartic potential}}, \href{https://doi.org/10.1140/epjc/s10052-024-12489-z}{\emph{Eur. Phys. J. C} {\bfseries 84} (2024) 127} [\href{https://arxiv.org/abs/2310.11030}{{\ttfamily 2310.11030}}].

\bibitem{Heydari:2024bxj}
S.~Heydari and K.~Karami, \emph{{Primordial black holes generated by fast-roll mechanism in non-canonical natural inflation}},  \href{https://arxiv.org/abs/2405.08563}{{\ttfamily 2405.08563}}.

\bibitem{Ballesteros:2018wlw}
G.~Ballesteros, J.~Beltran~Jimenez and M.~Pieroni, \emph{{Black hole formation from a general quadratic action for inflationary primordial fluctuations}}, \href{https://doi.org/10.1088/1475-7516/2019/06/016}{\emph{JCAP} {\bfseries 06} (2019) 016} [\href{https://arxiv.org/abs/1811.03065}{{\ttfamily 1811.03065}}].

\bibitem{Ballesteros:2021fsp}
G.~Ballesteros, S.~C\'espedes and L.~Santoni, \emph{{Large power spectrum and primordial black holes in the effective theory of inflation}}, \href{https://doi.org/10.1007/JHEP01(2022)074}{\emph{JHEP} {\bfseries 01} (2022) 074} [\href{https://arxiv.org/abs/2109.00567}{{\ttfamily 2109.00567}}].

\bibitem{Ashoorioon:2019xqc}
A.~Ashoorioon, A.~Rostami and J.T.~Firouzjaee, \emph{{EFT compatible PBHs: effective spawning of the seeds for primordial black holes during inflation}}, \href{https://doi.org/10.1007/JHEP07(2021)087}{\emph{JHEP} {\bfseries 07} (2021) 087} [\href{https://arxiv.org/abs/1912.13326}{{\ttfamily 1912.13326}}].

\bibitem{Frolovsky:2022ewg}
D.~Frolovsky, S.V.~Ketov and S.~Saburov, \emph{{Formation of primordial black holes after Starobinsky inflation}}, \href{https://doi.org/10.1142/S0217732322501358}{\emph{Mod. Phys. Lett. A} {\bfseries 37} (2022) 2250135} [\href{https://arxiv.org/abs/2205.00603}{{\ttfamily 2205.00603}}].

\bibitem{Fu:2019ttf}
C.~Fu, P.~Wu and H.~Yu, \emph{{Primordial Black Holes from Inflation with Nonminimal Derivative Coupling}}, \href{https://doi.org/10.1103/PhysRevD.100.063532}{\emph{Phys. Rev. D} {\bfseries 100} (2019) 063532} [\href{https://arxiv.org/abs/1907.05042}{{\ttfamily 1907.05042}}].

\bibitem{Heydari:2021gea}
S.~Heydari and K.~Karami, \emph{{Primordial black holes in nonminimal derivative coupling inflation with quartic potential and reheating consideration}}, \href{https://doi.org/10.1140/epjc/s10052-022-10036-2}{\emph{Eur. Phys. J. C} {\bfseries 82} (2022) 83} [\href{https://arxiv.org/abs/2107.10550}{{\ttfamily 2107.10550}}].

\bibitem{Heydari:2021qsr}
S.~Heydari and K.~Karami, \emph{{Primordial black holes ensued from exponential potential and coupling parameter in nonminimal derivative inflation model}}, \href{https://doi.org/10.1088/1475-7516/2022/03/033}{\emph{JCAP} {\bfseries 03} (2022) 033} [\href{https://arxiv.org/abs/2111.00494}{{\ttfamily 2111.00494}}].

\bibitem{Kawai:2021edk}
S.~Kawai and J.~Kim, \emph{{Primordial black holes from Gauss-Bonnet-corrected single field inflation}}, \href{https://doi.org/10.1103/PhysRevD.104.083545}{\emph{Phys. Rev. D} {\bfseries 104} (2021) 083545} [\href{https://arxiv.org/abs/2108.01340}{{\ttfamily 2108.01340}}].

\bibitem{Kawaguchi:2022nku}
R.~Kawaguchi and S.~Tsujikawa, \emph{{Primordial black holes from Higgs inflation with a Gauss-Bonnet coupling}}, \href{https://doi.org/10.1103/PhysRevD.107.063508}{\emph{Phys. Rev. D} {\bfseries 107} (2023) 063508} [\href{https://arxiv.org/abs/2211.13364}{{\ttfamily 2211.13364}}].

\bibitem{Ozsoy:2020kat}
O.~\"Ozsoy and Z.~Lalak, \emph{{Primordial black holes as dark matter and gravitational waves from bumpy axion inflation}}, \href{https://doi.org/10.1088/1475-7516/2021/01/040}{\emph{JCAP} {\bfseries 01} (2021) 040} [\href{https://arxiv.org/abs/2008.07549}{{\ttfamily 2008.07549}}].

\bibitem{Chen:2006nt}
X.~Chen, M.-x.~Huang, S.~Kachru and G.~Shiu, \emph{{Observational signatures and non-Gaussianities of general single field inflation}}, \href{https://doi.org/10.1088/1475-7516/2007/01/002}{\emph{JCAP} {\bfseries 01} (2007) 002} [\href{https://arxiv.org/abs/hep-th/0605045}{{\ttfamily hep-th/0605045}}].

\bibitem{Seery:2005wm}
D.~Seery and J.E.~Lidsey, \emph{{Primordial non-Gaussianities in single field inflation}}, \href{https://doi.org/10.1088/1475-7516/2005/06/003}{\emph{JCAP} {\bfseries 06} (2005) 003} [\href{https://arxiv.org/abs/astro-ph/0503692}{{\ttfamily astro-ph/0503692}}].

\bibitem{Riotto:2023hoz}
A.~Riotto, \emph{{The Primordial Black Hole Formation from Single-Field Inflation is Not Ruled Out}},  \href{https://arxiv.org/abs/2301.00599}{{\ttfamily 2301.00599}}.

\bibitem{Riotto:2023gpm}
A.~Riotto, \emph{{The Primordial Black Hole Formation from Single-Field Inflation is Still Not Ruled Out}},  \href{https://arxiv.org/abs/2303.01727}{{\ttfamily 2303.01727}}.

\bibitem{Choudhury:2023vuj}
S.~Choudhury, M.R.~Gangopadhyay and M.~Sami, \emph{{No-go for the formation of heavy mass Primordial Black Holes in Single Field Inflation}},  \href{https://arxiv.org/abs/2301.10000}{{\ttfamily 2301.10000}}.

\bibitem{Choudhury:2023jlt}
S.~Choudhury, S.~Panda and M.~Sami, \emph{{PBH formation in EFT of single field inflation with sharp transition}}, \href{https://doi.org/10.1016/j.physletb.2023.138123}{\emph{Phys. Lett. B} {\bfseries 845} (2023) 138123} [\href{https://arxiv.org/abs/2302.05655}{{\ttfamily 2302.05655}}].

\bibitem{Choudhury:2023rks}
S.~Choudhury, S.~Panda and M.~Sami, \emph{{Quantum loop effects on the power spectrum and constraints on primordial black holes}}, \href{https://doi.org/10.1088/1475-7516/2023/11/066}{\emph{JCAP} {\bfseries 11} (2023) 066} [\href{https://arxiv.org/abs/2303.06066}{{\ttfamily 2303.06066}}].

\bibitem{Franciolini:2023lgy}
G.~Franciolini, A.~Iovino, Junior., M.~Taoso and A.~Urbano, \emph{{One loop to rule them all: Perturbativity in the presence of ultra slow-roll dynamics}},  \href{https://arxiv.org/abs/2305.03491}{{\ttfamily 2305.03491}}.

\bibitem{Davies:2023hhn}
M.W.~Davies, L.~Iacconi and D.J.~Mulryne, \emph{{Numerical 1-loop correction from a potential yielding ultra-slow-roll dynamics}}, \href{https://doi.org/10.1088/1475-7516/2024/04/050}{\emph{JCAP} {\bfseries 04} (2024) 050} [\href{https://arxiv.org/abs/2312.05694}{{\ttfamily 2312.05694}}].

\bibitem{Fumagalli:2023loc}
J.~Fumagalli, S.~Bhattacharya, M.~Peloso, S.~Renaux-Petel and L.T.~Witkowski, \emph{{One-loop infrared rescattering by enhanced scalar fluctuations during inflation}}, \href{https://doi.org/10.1088/1475-7516/2024/04/029}{\emph{JCAP} {\bfseries 04} (2024) 029} [\href{https://arxiv.org/abs/2307.08358}{{\ttfamily 2307.08358}}].

\bibitem{Inomata:2022yte}
K.~Inomata, M.~Braglia, X.~Chen and S.~Renaux-Petel, \emph{{Questions on calculation of primordial power spectrum with large spikes: the resonance model case}}, \href{https://doi.org/10.1088/1475-7516/2023/04/011}{\emph{JCAP} {\bfseries 04} (2023) 011} [\href{https://arxiv.org/abs/2211.02586}{{\ttfamily 2211.02586}}].

\bibitem{Caravano:2024tlp}
A.~Caravano, K.~Inomata and S.~Renaux-Petel, \emph{{The Inflationary Butterfly Effect: Non-Perturbative Dynamics From Small-Scale Features}},  \href{https://arxiv.org/abs/2403.12811}{{\ttfamily 2403.12811}}.

\bibitem{Mishra:2023lhe}
S.S.~Mishra, E.J.~Copeland and A.M.~Green, \emph{{Primordial black holes and stochastic inflation beyond slow roll. Part I. Noise matrix elements}}, \href{https://doi.org/10.1088/1475-7516/2023/09/005}{\emph{JCAP} {\bfseries 09} (2023) 005} [\href{https://arxiv.org/abs/2303.17375}{{\ttfamily 2303.17375}}].

\bibitem{Jackson:2023obv}
J.H.P.~Jackson, H.~Assadullahi, A.D.~Gow, K.~Koyama, V.~Vennin and D.~Wands, \emph{{The separate-universe approach and sudden transitions during inflation}},  \href{https://arxiv.org/abs/2311.03281}{{\ttfamily 2311.03281}}.

\bibitem{Firouzjahi:2023ahg}
H.~Firouzjahi and A.~Riotto, \emph{{Primordial Black Holes and loops in single-field inflation}}, \href{https://doi.org/10.1088/1475-7516/2024/02/021}{\emph{JCAP} {\bfseries 02} (2024) 021} [\href{https://arxiv.org/abs/2304.07801}{{\ttfamily 2304.07801}}].

\bibitem{Iacconi:2023ggt}
L.~Iacconi, D.~Mulryne and D.~Seery, \emph{{Loop corrections in the separate universe picture}},  \href{https://arxiv.org/abs/2312.12424}{{\ttfamily 2312.12424}}.

\bibitem{Motohashi:2023syh}
H.~Motohashi and Y.~Tada, \emph{{Squeezed bispectrum and one-loop corrections in transient constant-roll inflation}}, \href{https://doi.org/10.1088/1475-7516/2023/08/069}{\emph{JCAP} {\bfseries 08} (2023) 069} [\href{https://arxiv.org/abs/2303.16035}{{\ttfamily 2303.16035}}].

\bibitem{Tasinato:2023ioq}
G.~Tasinato, \emph{{Non-Gaussianities and the large |\ensuremath{\eta}| approach to inflation}}, \href{https://doi.org/10.1103/PhysRevD.109.063510}{\emph{Phys. Rev. D} {\bfseries 109} (2024) 063510} [\href{https://arxiv.org/abs/2312.03498}{{\ttfamily 2312.03498}}].

\bibitem{Tasinato:2023ukp}
G.~Tasinato, \emph{{Large |\ensuremath{\eta}| approach to single field inflation}}, \href{https://doi.org/10.1103/PhysRevD.108.043526}{\emph{Phys. Rev. D} {\bfseries 108} (2023) 043526} [\href{https://arxiv.org/abs/2305.11568}{{\ttfamily 2305.11568}}].

\bibitem{Jarnhus:2007ia}
P.R.~Jarnhus and M.S.~Sloth, \emph{{de Sitter limit of inflation and nonlinear perturbation theory}}, \href{https://doi.org/10.1088/1475-7516/2008/02/013}{\emph{JCAP} {\bfseries 02} (2008) 013} [\href{https://arxiv.org/abs/0709.2708}{{\ttfamily 0709.2708}}].

\bibitem{Arroja:2008ga}
F.~Arroja and K.~Koyama, \emph{{Non-gaussianity from the trispectrum in general single field inflation}}, \href{https://doi.org/10.1103/PhysRevD.77.083517}{\emph{Phys. Rev. D} {\bfseries 77} (2008) 083517} [\href{https://arxiv.org/abs/0802.1167}{{\ttfamily 0802.1167}}].

\bibitem{Maity:2023qzw}
S.~Maity, H.V.~Ragavendra, S.K.~Sethi and L.~Sriramkumar, \emph{{Loop contributions to the scalar power spectrum due to quartic order action in ultra slow roll inflation}},  \href{https://arxiv.org/abs/2307.13636}{{\ttfamily 2307.13636}}.

\bibitem{Akhshik:2015nfa}
M.~Akhshik, H.~Firouzjahi and S.~Jazayeri, \emph{{Effective Field Theory of non-Attractor Inflation}}, \href{https://doi.org/10.1088/1475-7516/2015/07/048}{\emph{JCAP} {\bfseries 07} (2015) 048} [\href{https://arxiv.org/abs/1501.01099}{{\ttfamily 1501.01099}}].

\bibitem{Firouzjahi:2023aum}
H.~Firouzjahi, \emph{{One-loop corrections in power spectrum in single field inflation}}, \href{https://doi.org/10.1088/1475-7516/2023/10/006}{\emph{JCAP} {\bfseries 10} (2023) 006} [\href{https://arxiv.org/abs/2303.12025}{{\ttfamily 2303.12025}}].

\bibitem{Fumagalli:2023hpa}
J.~Fumagalli, \emph{{Absence of one-loop effects on large scales from small scales in non-slow-roll dynamics}},  \href{https://arxiv.org/abs/2305.19263}{{\ttfamily 2305.19263}}.

\bibitem{Tada:2023rgp}
Y.~Tada, T.~Terada and J.~Tokuda, \emph{{Cancellation of quantum corrections on the soft curvature perturbations}}, \href{https://doi.org/10.1007/JHEP01(2024)105}{\emph{JHEP} {\bfseries 01} (2024) 105} [\href{https://arxiv.org/abs/2308.04732}{{\ttfamily 2308.04732}}].

\bibitem{Firouzjahi:2023bkt}
H.~Firouzjahi, \emph{{Revisiting loop corrections in single field ultraslow-roll inflation}}, \href{https://doi.org/10.1103/PhysRevD.109.043514}{\emph{Phys. Rev. D} {\bfseries 109} (2024) 043514} [\href{https://arxiv.org/abs/2311.04080}{{\ttfamily 2311.04080}}].

\bibitem{Braglia:2024zsl}
M.~Braglia and L.~Pinol, \emph{{No time to derive: unraveling total time derivatives in in-in perturbation theory}},  \href{https://arxiv.org/abs/2403.14558}{{\ttfamily 2403.14558}}.

\bibitem{Kawaguchi:2024lsw}
R.~Kawaguchi, S.~Tsujikawa and Y.~Yamada, \emph{{Roles of boundary and equation-of-motion terms in cosmological correlation functions}},  \href{https://arxiv.org/abs/2403.16022}{{\ttfamily 2403.16022}}.

\bibitem{Firouzjahi:2024psd}
H.~Firouzjahi, \emph{{Loop Corrections in Bispectrum in USR Inflation with PBHs Formation}},  \href{https://arxiv.org/abs/2403.03841}{{\ttfamily 2403.03841}}.

\bibitem{Syu:2019uwx}
W.-C.~Syu, D.-S.~Lee and K.-W.~Ng, \emph{{Quantum loop effects to the power spectrum of primordial perturbations during ultra slow-roll inflation}}, \href{https://doi.org/10.1103/PhysRevD.101.025013}{\emph{Phys. Rev. D} {\bfseries 101} (2020) 025013} [\href{https://arxiv.org/abs/1907.13089}{{\ttfamily 1907.13089}}].

\bibitem{Cheng:2021lif}
S.-L.~Cheng, D.-S.~Lee and K.-W.~Ng, \emph{{Power spectrum of primordial perturbations during ultra-slow-roll inflation with back reaction effects}}, \href{https://doi.org/10.1016/j.physletb.2022.136956}{\emph{Phys. Lett. B} {\bfseries 827} (2022) 136956} [\href{https://arxiv.org/abs/2106.09275}{{\ttfamily 2106.09275}}].

\bibitem{Cheng:2023ikq}
S.-L.~Cheng, D.-S.~Lee and K.-W.~Ng, \emph{{Primordial perturbations from ultra-slow-roll single-field inflation with quantum loop effects}}, \href{https://doi.org/10.1088/1475-7516/2024/03/008}{\emph{JCAP} {\bfseries 03} (2024) 008} [\href{https://arxiv.org/abs/2305.16810}{{\ttfamily 2305.16810}}].

\bibitem{Saburov:2024und}
S.~Saburov and S.V.~Ketov, \emph{{Quantum loop corrections in the modified gravity model of Starobinsky inflation with primordial black hole production}},  \href{https://arxiv.org/abs/2402.02934}{{\ttfamily 2402.02934}}.

\bibitem{Ballesteros:2024zdp}
G.~Ballesteros and J.G.~Egea, \emph{{One-loop power spectrum in ultra slow-roll inflation and implications for primordial black hole dark matter}},  \href{https://arxiv.org/abs/2404.07196}{{\ttfamily 2404.07196}}.

\bibitem{Inomata:2024lud}
K.~Inomata, \emph{{Curvature Perturbations Protected Against One Loop}},  \href{https://arxiv.org/abs/2403.04682}{{\ttfamily 2403.04682}}.

\end{thebibliography}\endgroup

\end{document}